\documentclass{aa}  
\usepackage{graphicx}
\usepackage{xcolor}
\usepackage{comment}
\usepackage{amsmath}
\usepackage{txfonts}
\usepackage{ulem}

\usepackage{hyperref}
\hypersetup{
    colorlinks = true,
    citecolor = orange,
}

\makeatletter
\renewcommand*\aa@pageof{, page \thepage{} of \pageref*{LastPage}}
\makeatother

\newcommand{\alphac}{\alpha_\mathrm{c}}
\newcommand{\alphap}{\alpha_\mathrm{p}}
\newcommand{\betac}{\beta_\mathrm{c}}
\newcommand{\betap}{\beta_\mathrm{p}}

\newcommand{\kp}{k_\mathrm{p}}
\newcommand{\Xc}{X_\mathrm{c}}
\newcommand{\Xp}{X_\mathrm{p}}
\newcommand{\mc}{m_\mathrm{c}}
\newcommand{\massp}{m_\mathrm{p}}
\newcommand{\mtot}{m_\mathrm{tot}}

\definecolor{goldenpoppy}{rgb}{0.99, 0.76, 0.0}

\begin{document} 

\title{A new pulsar timing model for scalar-tensor gravity}
\subtitle{with applications to PSR~J2222$-$0137 and pulsar-black hole binaries}


\author{
  A.~Batrakov\inst{1} \and
  H.~Hu\inst{1} \and
  N.~Wex\inst{1} \and
  P.~C.~C. Freire\inst{1} \and
  V.~Venkatraman Krishnan\inst{1} \and
  M.~Kramer\inst{1,2} \and \\ 
  Y.~J.~Guo\inst{1} \and 
  L.~Guillemot\inst{3,4} \and
  J.~W.~McKee\inst{5,6,7} \and 
  I.~Cognard\inst{3,4} \and  
  G.~Theureau\inst{3,4}
}

\institute{
  Max-Planck-Institut f\"ur Radioastronomie, Auf dem H\"ugel 69, 53121 Bonn, Germany \\
  \email{abatrakov@mpifr-bonn.mpg.de}
  \and
  Jodrell Bank Centre for Astrophysics, The University of Manchester, M13 9PL, United Kingdom
  \and
  Observatoire Radioastronomique de Nan{\c c}ay, Observatoire de Paris, Universit\'e PSL, Universit\'e d’Orl\'eans, CNRS, 18330 Nan{\c c}ay, France
  \and
  Laboratoire de Physique et Chimie de l'Environnement et de l'Espace, Universit\'e d’Orl\'eans/CNRS, 45071 Orl\'eans Cedex 02, France
  \and
  Canadian Institute for Theoretical Astrophysics, University of Toronto, 60 St. George Street, Toronto, ON M5S 3H8, Canada
  \and
  E.A. Milne Centre for Astrophysics, University of Hull, Cottingham Road, Kingston-upon-Hull, HU6 7RX, UK
  \and
  Centre of Excellence for Data Science, Artificial Intelligence and Modelling (DAIM), University of Hull, Cottingham Road, Kingston-upon-Hull, HU6 7RX, UK
}

\date{Received MM DD, YYYY; accepted MM DD, YYYY}

 
\abstract
  {Scalar-tensor gravity (STG) theories are well-motivated alternatives to general relativity (GR). One class of STG theories, the Damour-Esposito-Far{\`e}se (DEF) gravity, has a massless scalar field with two arbitrary coupling parameters. We are interested in this theory because, despite its simplicity, it predicts a wealth of different phenomena, such as dipolar gravitational wave emission and spontaneous scalarisation of neutron stars (NSs). These phenomena of DEF gravity can be tested by timing binary radio pulsars. In methods used so far, intermediate phenomenological post-Keplerian (PK) parameters are measured by fitting the corresponding timing model to the timing data whose values are then compared to the predictions from the alternative theory under test. However, this approach loses information between intermediate steps and does not account for possible correlations between PK parameters.}
  {We aim to develop a new binary pulsar timing model DDSTG to enable more precise tests of STG theories based on a minimal set of binary parameters. The expressions for PK parameters in DEF gravity are self-consistently incorporated into the model. PK parameters depend on two masses which are now directly fitted to the data without intermediate steps. The new technique takes into account all possible correlations between PK parameters naturally.}
  {Grids of physical parameters of NSs are calculated in the framework of DEF gravity for a set of 11 equations of state. The automatic Differentiation (AutoDiff) technique is employed, which aids in the calculation of gravitational form factors of NSs with higher precision than in previous works. The pulsar timing program \textsc{TEMPO} is selected as a framework for the realisation of the DDSTG model. The implemented model is applicable to any type of pulsar companions. We also simulate realistic future radio-timing data-sets for a number of large radio observatories for the binary pulsar PSR~J2222$-$0137 and three generic pulsar-black hole (PSR-BH) systems.}
  {We apply the DDSTG model to the most recently published observational data for PSR~J2222$-$0137. The obtained limits on DEF gravity parameters for this system confirm and improve previous results. New limits are also the most reliable because DEF gravity is directly fitted to the data. We argue that future observations of PSR~J2222$-$0137 can significantly improve the limits and that PSR-BH systems have the potential to place the tightest limits in certain areas of the DEF gravity parameter space.}
  {}


\maketitle

\section{Introduction}
\label{sec:introduction}

General Relativity (GR) proved to be the most successful theory of gravity for more than a century. Up to now, it has passed all experimental tests with flying colours. The weak field regime is verified by the Solar System experiments \citep{wil14}, whereas strong field effects are especially well tested by the timing of binary pulsars \citep{wk20}. Among other things, pulsar tests enable precise tests of the radiative properties of gravity and the strong equivalence principle (SEP). Furthermore, the large-scale behaviour of gravity (low spacetime curvature and temporal variation) is tested in cosmological observations \citep{cfp12}. Finally, in the last few years,  ground-based gravitational wave (GW) detectors observed GWs from coalescing binary black holes \citep{ligo16, ligo17, ligo19, ligo21}, binary neutron stars \citep{ligo17b, ligo17c, ligo19b} and also black hole-neutron star binaries \citep{ligo21b}; probing the hitherto unexplored highly dynamic, strong-field regime of gravity. Thus far, GR can account for all observed effects both in weak and strong gravitational fields, in the quasi-static and the highly dynamical regime, and on small as well as on large scales.\footnote{Within GR, the large scale requires the introduction of dark matter and dark energy (latter in form of the cosmological constant $\Lambda$).}

Despite such successes, there are still convincing reasons to investigate modified theories of gravity \citep{bbc+15}. GR describes gravitational interaction by a single massless, spin-2 tensor field $g_{\mu\nu}$, known as the metric of spacetime. Some of the best motivated alternative gravity theories, the scalar-tensor gravity (STG) theories,  incorporate additional scalar degrees of freedom to mediate gravity. Such scalar fields arise naturally in higher dimensional theories in their low energy limits, e.g., the old Kaluza-Klein theory \citep{kal21,kle26} and
string theories \citep{fm07}. Scalar-tensor gravity theories are also motivated by cosmological questions of inflation, dark energy and even the thus far undiscovered quantum gravity \citep{cfp+12}. 

The STG theory that for many years was considered as the only natural competitor of GR is known as Jordan-Fierz-Brans-Dicke (JFBD) gravity \citep{jor55, jor59, fie56, bd61}.\footnote{The first to formulate JFBD gravity was actually Willy Scherrer in the early 1940s (see \cite{goe12} for a historical review on the genesis of STG).} JFBD gravity is a metric theory (matter and non-gravitational fields couple only to one specific spacetime metric) and therefore fulfils the Einstein equivalence principle by design \citep{wil93,dam12}. However, an additional scalar field is nonminimally coupled to matter via the choice of a special conformal coupling function; thus, the strong equivalence principle (SEP) is violated \citep{wil14}. This coupling function depends on a single parameter $\omega_{\mathrm{BD}}$, which by now is tightly constrained, in particular with the pulsar in the stellar triple system \citep{vcf+20}. \cite{ber68} and \cite{wag70} presented the most general STG with one scalar field which is in its action at most quadratic in the derivatives of the fields. The most general mono-scalar–tensor theory with second-order field equations is Horndeski gravity \citep{hor74}. In 1992, \cite{de92a} presented a generic class of STG theories with an arbitrary number of scalar fields. In the scope of this paper, we mainly focus on {\it DEF gravity} \citep{de93new}. DEF gravity is a Bergman-Wagoner theory which has a massless scalar field and a quadratic coupling function with two arbitrary parameters. 

Apart from being well-motivated, DEF gravity predicts interesting effects which are not present in weak fields of the Solar System but could be tested by employing pulsar astronomy. Neutron stars (NSs) have strong gravitational fields and large gravitational binding energies due to their high compactness. Because of this, in STG theories they can have large gravitational form-factors (also known as scalar charges), unlike weakly self-gravitating objects like normal stars and white dwarfs (WDs). Consequently, while STG theories generally predict the emission of dipolar and monopolar GWs, which increase the rate of orbital period decay relative to the GR expectation; this is expected to be especially strong in asymmetric binary systems containing a NS and a WD, due to the significant differences in their compactness and scalar charges.

Furthermore, \cite{de93new, de96new} found a specific phenomenon in NSs called ``spontaneous scalarisation''. This fully non-perturbative effect excites the scalar field above the cosmological background value in NSs, allowing scalar charges of order unity, even if there is no deviation from GR in the weak-field regime. This could produce, for binary systems containing NSs with specific masses, a highly enhanced rate of dipolar GW emission.

Some of the tightest tests of gravity with strongly self-gravitating objects are provided by the high precision timing of binary pulsars \citep{stai03,wex14,shw16,wk20,ksm21}. These tests are performed in a quasi-stationary strong-field gravity regime, where the gravitational field is strong near and inside NSs and produces high curvature, whereas the velocities are small compared to the speed of light $v/c \sim 10^{-3}$. The first timing model (BT) was introduced by \cite{bt76} and allowed to extract information from the timing of binary pulsars. An extended model covering all relativistic effects in the dynamics of a binary system up to the first post-Newtonian level was proposed later by \cite{dd85, dd86} (DD).

The DD model gave an opportunity to perform self-consistent tests of gravity by means of the parametrised post-Keplerian (PPK) formalism (see, for instance, applications by \citealt{tw89}). In this generic framework, theory-independent Keplerian and post-Keplerian (PK) timing parameters, which quantify the relativistic motion of the pulsar and the propagation of its radio signals, are fitted to the observational timing data. Apart from the well measured Keplerian parameters of the orbit, each PK parameter depends only on the masses of the pulsar and its companion. As soon as the expressions for the PK parameters in a particular gravity theory are known, these two masses can be derived from the PK parameters once at least two of them are known. If more than two PK parameters become available, then a test of the consistency of that gravity theory can be made. Later, \cite{dt92} expanded the DD model and introduced pulse structure parameters. They also showed that the DD model can be applied to a large set of fully-conservative theories of gravity and presented phenomenological PK expressions in a framework of generic boost-invariant theories (the modified Einstein-Infeld-Hoffmann formalism).

However, such an independent measurement of PK parameters means that any correlations between them inadvertently reduce the sensitivity of their measurements. \cite{tw89} solved this problem for GR with the introduction of the DDGR model, where the GR expressions for PK parameters are incorporated in the model, and two masses are directly fitted to the observational timing data. This mitigates or breaks any correlations between the PK parameters.

In this work we present a new pulsar timing model, the ``DDSTG'' model. The idea is very similar to that of the DDGR model: fitting directly for the masses of the components from pulsar timing data, but instead of GR using a particular member of the two-parametric DEF gravity. Apart from the theoretical predictions for all PK parameters within that theory, the model uses pre-calculated grids of physical parameters (gravitational form-factors) of NSs in that particular gravity theory. The direct fit of two masses of the companions reduces the model to a minimal set of parameters and naturally solves all the issues of possible correlations between the observed parameters. This feature, achieved by the construction, makes the DDSTG model superior to the traditional methods, e.g.\ the ``PK method'' based on the measurement of PK parameters.

As an application of the new timing model DDSTG, we use the most recently published timing data for PSR~J2222$-$0137 from \cite{cfg+17} and \cite{gfg21}. This pulsar is interesting for several reasons: It is one of the closest pulsars known and has a good timing precision. A precise measurement of the variation of the orbital period ($\dot{P}_\mathrm{b}$) and the high asymmetry in compactness between the pulsar and its companion allows tight constraints on the emission of dipolar GWs. Moreover, the pulsar's mass $m_{\mathrm{p}} = 1.831(10) M_{\odot}$ \citep{gfg21} lies in the range ($\massp \gtrsim 1.5 M_{\odot}$) subject to the spontaneous scalarisation effect; the non-detection of dipolar GW emission in this system has introduced strong constraints on the existence of that phenomenon, at least within the DEF framework \citep{zfk+22}. 

We organise the paper as follows: in Section~\ref{sec:DEF}, we introduce the basics of the Damour-Esposito-Far{\`e}se (DEF) gravity and present our machinery preparations necessary for the implementation of the new model. In Section~\ref{sec:DDSTG}, we discuss so far used standard timing models and introduce our new model DDSTG for testing STG theories. The next sections are devoted to the applications of our new timing model. In Section~\ref{sec:J2222_now}, we provide a brief description of the \cite{gfg21} data-set on PSR~J2222$-$0137 and apply the DDSTG model to it, discuss the limits on DEF gravity and demonstrate that our new method provides indeed tighter and more reliable constraints than previous methods. In Section~\ref{sec:J2222_future}, based on simulated data-sets for the largest radio observatories, we discuss what limits on these theories might be achieved in the near future with continued timing of PSR~J2222$-$0137 . In Section~\ref{sec:PSRBH}, we show possible future constraints from reasonable pulsar-black hole (PSR-BH) binary systems. We also briefly discuss the possible origin of PSR-BH systems and how it affects the test if such a system is located in a globular cluster (GC). Section~\ref{sec:other_theories} is devoted to a discussion of other alternative gravity theories and possibilities of DDSTG extensions beyond DEF gravity. In Section \ref{sec:gdot} we discuss the effect of the time-varying gravitational constant as a perspective future extension of the DDSTG model. In Section~\ref{sec:conclusions} we summarise our results.

\section{Damour-Esposito-Far{\`e}se gravity}
\label{sec:DEF}

This paper investigates STG theories, natural alternatives to GR. Specifically, we mainly focus on DEF gravity. In the following we summarise aspects of that class of gravity theories that are relevant for this paper.

\subsection{Theoretical aspects}

DEF gravity is a STG theory with one long range, massless scalar field $\varphi$ non-minimally coupled to the curvature scalar. The theory is fully described by the action, which is most simply presented in the {\it Einstein frame} \citep{de93new,de96new} 
\begin{equation}
    \label{eq:action}
    S = \dfrac{c^4}{16\pi G_*} \int \dfrac{d^4x}{c} g^{1/2}_* (R_* - 2g^{\mu \nu}_* \partial_{\mu} \varphi \partial_{\nu} \varphi) + S_m \left[\psi_m; A^2(\varphi) g^*_{\mu \nu}\right] \, ,
\end{equation}
where $G_*$ is the bare gravitational coupling constant, $g_*$ and $R_*$ are the determinant and the Ricci scalar associated with the Einstein metric ($g^*_{\mu \nu}$). The last term in Eq.~\eqref{eq:action} describes an action associated with any matter fields ($\psi_m$). $A(\varphi)$ is the coupling function and takes a specific exponential form in DEF gravity: $A(\varphi) = \exp[\alpha_0(\varphi - \varphi_0) + \frac12\beta_0 (\varphi - \varphi_0)^2]$. Introduced quantities $\{\alpha_0, \beta_0\}$ are two arbitrary parameters of the theory and $\varphi_0$ is the scalar field at spatial infinity. The coupling function $A(\varphi)$ plays the role of a conformal factor connecting the ``physical metric'' $\Tilde{g}_{\mu \nu} = A^2(\varphi)g^*_{\mu \nu}$ with the Einstein metric. The metric $\Tilde{g}_{\mu \nu}$ is the one measured by laboratory clocks and rods due to the universal coupling of matter to this metric. The bare gravitational constant is simply related to the one measured in a Cavendish experiment by $G_{\mathrm{Cav}} = G_* (1+\alpha_0^2)$. GR corresponds to vanishing coupling parameters, i.e.\ $\alpha_0 = \beta_0 = 0$, and JFBD gravity is recovered by setting $\beta_0 = 0$ while keeping $\alpha_0 \ne 0$.

The field equations are derived by variation of the action in Eq.~\eqref{eq:action} and are most simply formulated in terms of the pure-spin field variables $({g^*_{\mu \nu}}, \varphi)$ in the Einstein frame:
\begin{eqnarray}
    \label{eq:einstein}
    R^*_{\mu \nu} &=& 2\partial_{\mu} \varphi \partial_{\nu} \varphi + \dfrac{8\pi G_*}{c^4} \left(T^*_{\mu \nu} - \dfrac12 T^*g^*_{\mu \nu}\right),\\
    \Box_{g^*} \varphi &=& - \dfrac{8\pi G_*}{c^4} \alpha(\varphi) T_*,
\end{eqnarray}
with a material stress-energy tensor $T_*^{\mu \nu} = 2c g_*^{-1/2} \delta S_m / \delta g^*_{\mu \nu}$ and $\alpha(\varphi) = \partial \ln A(\varphi) / \partial \varphi$, which measures the coupling strength between the scalar field and matter.

All weak-field deviations from GR may be described in terms of the asymptotic behaviour of $\alpha(\varphi)$ at spatial infinity. The theory parameters $\alpha_0 = \alpha(\varphi_0),\ \beta_0 = \beta(\varphi_0)$ can be interpreted as asymptotic values of the function $\alpha(\varphi)$ and its derivative $\beta(\varphi) = \partial \alpha(\varphi)/\partial \varphi$.

In this work, without loss of generality, we use the formulation of DEF gravity with $\varphi_0 \equiv 0$, two parameters $\{\alpha_0, \beta_0\}$ appearing in the form of $A(\varphi)$ and $A(\varphi_0) =1$. There is an equivalent formulation in terms of parameters $\{\beta_0, \varphi_0\}$, where $A(\varphi) = \exp(\beta_0\varphi^2/2)$ and $\varphi_0 = \alpha_0/\beta_0$. But the latter formulation is not well determined for the JFBD limit $\beta_0=0$ \citep{dam07}.

The parameterised post-Newtonian (PPN) framework allows describing a wide range of metric theories of gravity in their weak field approximation by ten independent PPN parameters. DEF gravity is a fully conservative theory and has only two PPN parameters that deviate from their GR values, which are also called Eddington parameters \citep{wil93}:
\begin{eqnarray}
    \label{eq:eddington}
    \gamma_{\mathrm{Edd}} &=& 1 - \frac{2\alpha_0^2}{1 + \alpha_0^2} \,,\\
    \beta_{\mathrm{Edd}}  &=& 1 + \frac{\beta_0\alpha_0^2}{2(1 + \alpha_0^2)^2} \,.
\end{eqnarray}
The PPN parameter $\gamma_{\mathrm{Edd}}$ measures the spatial curvature induced by unit rest mass and $\beta_{\mathrm{Edd}}$ is a measure of the amount of nonlinearity in the superposition law of gravity. Solar System experiments put limits on these parameters. The $| \gamma_{\mathrm{Edd}} - 1| \lesssim 2.3 \times 10^{-5}$ limit comes from the Shapiro time delay observed by the Cassini spacecraft \citep{bit03, wil14}. The $| \beta_{\mathrm{Edd}} - 1| \lesssim 8 \times 10^{-5}$ limit comes from observations of the perihelion shift of Mercury \citep{wil14}. The Cassini experiment yields a direct limit on $\alpha_0^2 \lesssim 1.15\times10^{-5}$, whereas $\beta_0$ remains unconstrained from weak-field Solar System experiments, since $\alpha_0$ can be arbitrarily small.

\subsection{Scalarisation of neutron stars in DEF gravity}

To place firm limits on $\beta_0$ parameter, we need to explore the strong-field effects of DEF gravity. One can expect deviations from GR in the presence of highly compact massive objects, in particular for NSs, due to the  scalar field sourced by the strong internal gravitational field of the star. When placing limits on $\beta_0$ it is important that for certain areas in the DEF gravity parameter space one has a significant growth of the scalar field in the NS interior, even if the deviations for weakly self-gravitating bodies are very small, or even zero \citep{de92,de93new,de96new}. The origin of this spontaneous scalarisation lies in the tachyonic instability of the field equations happening when a NS reaches a sufficiently high compactness. The mechanism of this instability is similar to what happens in ferromagnets during spontaneous magnetisation \citep{de96new}. It is a fully nonlinear non-perturbative effect, thus it requires accurate techniques to be properly analysed.

For testing DEF gravity, it is crucial to introduce three gravitational form-factors appearing in the expressions for PK parameters
\begin{equation}
    \label{eq:scalar_charges}
    \alpha_A = \dfrac{\partial \ln m_A}{\partial \varphi_{0}} \Big|_{\Bar{m}_A},\quad \beta_A = \dfrac{\partial \alpha_A}{\partial \varphi_{0}} \Big|_{\Bar{m}_A},\quad k_A = -\dfrac{\partial \ln I_A}{\partial \varphi_{0}} \Big|_{\Bar{m}_A} \,,
\end{equation}
where all the derivatives are taken for a fixed baryonic mass ($\Bar{m}_A$).
The important quantity $\alpha_A$ counts the effective couple strength between the ambient scalar field and a NS. The subscript A refers to the star A. $\alpha_A$ is a strong-field counterpart to the parameter $\alpha_0$, which is its weak field approximation. If a NS mass exceeds the critical mass of $m_\mathrm{crit}$ then $\alpha_{\mathrm{NS}}$ can become suddenly of order of unity $\alpha_{\mathrm{NS}} \sim O(1)$ due to the scalarisation effect, while its weak field counterpart stays close to zero $\alpha_0 \sim 0$. The value of the critical mass depends on $\beta_0$ and the chosen equation of state (EOS). The second parameter $\beta_A$ reflects a change of the scalar charge when the asymptotic scalar field changes. The derivative $\beta_A$ can also obtain very large values in the transition scalarisation region, where $\alpha_A$ changes fast with the change of a mass. The last quantity $k_A$ describes the field dependence of the moment of inertia $I_A$ and can become important in the eccentric systems.

\subsection{Calculating grids of NSs}\label{subsec:calc_grids}

To test DEF gravity in precision timing experiments of binary pulsars, one must know the exact properties of NSs in that specific class of alternative gravity. In our work, we follow the procedure described in \citet{de96new} for calculating NSs in DEF gravity. We assume the slowly rotating NS approximation with a stationary axisymmetric metric. Assuming this axial symmetry and neglecting terms of rotational velocity squared (or higher), the Einstein equations can be written as a system of 8 first-order ordinary differential equations (ODEs) for a set of radial functions (see the representation by \citet{de96new}). The equations are complemented by appropriate boundary conditions placed at spatial infinity and the NS centre. Boundary conditions at spatial infinity and thus properties of NSs depend on the values of the theory parameters $\{\alpha_0, \beta_0\}$.

For solving NS structures, we developed a modular program in the \textsc{Julia} language \citep{Julia-2017}. The program enables us to calculate both single NSs for selected parameters and grids of NSs for many desired parameters. The calculations are performed accurately and efficiently using the methods of parallel programming. Currently, DEF gravity is implemented as the chosen theory of gravity, but our program allows to use any coupling function $A(\varphi)$ to extend our tests beyond DEF gravity parametrisation.

The internal structure and NSs parameters significantly depend on EOS. The strong dependence of the scalarisation phenomenon on the choice of EOS was extensively analysed by \cite{ssb+17} and \cite{zfk+22} in order to put constraints on this non-linear phenomenon from radio pulsars. In our work, we use a piece-wise polytropic approximation for the EOSs following the procedure in \citet{rea09}. We select 11 EOSs that allow us to cover a range from soft to stiff while still being in agreement with the maximum mass observed for a NS (for more information about selected EOS we refer to Appendix~\ref{App:pre-calculated_grids}). For tests in the present paper, we select MPA1 high-density EOS \citep{mpa87} in its piece-wise polytropic form \citep{rea09} with three polytropic pieces. MPA1 is a stiff EOS with a maximum mass of a NS equal to $2.46 \, M_{\odot}$. For low densities, we use an analytic form of the SLy EOS proposed by \citet{dh01} and approximated by four polytropic pieces \citep{rea09}. In general stiff EOSs (e.g. MPA1) produce more conservative limits in most parts of DEF gravity parameter space.

In our program, we utilise the Automatic Differentiation (AutoDiff) technique (see e.g. \citet{mar18}) for calculating derivatives of NSs quantities. AutoDiff is a powerful alternative to numerical differentiation deprived of numerical errors (for detailed explanation see Appendix~\ref{App:AutoDiff}). AutoDiff algorithm utilises exact expressions of derivatives for elementary functions and a chain rule to calculate complex derivatives with the working machine-precision.  Thus our program with applied AutoDiff technique calculates gravitational form-factors $\{\alpha_A, \beta_A, k_A\}$, presented as derivatives in Eq.~\eqref{eq:scalar_charges}, with very high numerical precision.

Using the developed program, we calculate gravitational form-factors and masses for an extensive grid of DEF parameters and central pressures $\{\alpha_0, \beta_0, p_c\}$ assuming a specific EOS. Our grids cover ranges of $\alpha_0 \in [-10^{-1}, -10^{-5}],\ \beta_0 \in [-6, +10],\ p_c \in [10^{34}, 10^{36}]$ dyn/cm$^2$ and are sampled linearly for $\log\alpha_0, \log p_c$ and $\beta_0$. The size of each grid is $\{101,351,121\}$ points respectively. More technical details about the procedure of calculating NS properties, the AutoDiff technique, and the structure of pre-calculated grids can be found in Appendix~\ref{App:details_on_grids}. We stick to pre-calculating grids because it is numerically costly to calculate NS structures for a particular mass on the fly. Further one can perform interpolation over these pre-calculated grids or methods beyond interpolation, e.g reduced-order models in \citet{gzs21b}. The achieved accuracy of grid values and further interpolation over the grid is sufficient for our purposes.

The accuracy of the calculations can be checked by comparison of the scalar charge ($\alpha_A$) calculated as the derivative from Eq.~\eqref{eq:scalar_charges} $\alpha_{A,\,\mathrm{deriv}}$ and its direct value from the asymptotic behaviour of the external scalar field obtained after integration $\alpha_{A,\,\mathrm{asympt}}$ \citep{de93new}. The relative accuracy expressed by $\delta_{\mathrm{rel}} = |1 - \alpha_{A,\,\mathrm{deriv}} / \alpha_{A,\,\mathrm{asympt}}|$ lies in the range $\delta_{\mathrm{rel}} \sim 10^{-7}-10^{-13}$ for our calculations. The calculated scalar charges are significantly more accurate compared to previous works, e.g., \citet{and19} with typical accuracy of $\delta_{\mathrm{rel}} \sim 10^{-1}-10^{-5}$. However, it is essential to mention that the accuracy of several percent is already sufficient for testing DEF gravity with binary pulsar timing.

Besides, we also find that the NS structure appears to be very sensitive to internal thermodynamic consistency. The piece-wise polytropic EOSs fulfil the thermodynamic consistency by their definition. However, often one uses tabulated EOSs instead, which must be interpolated. Unfortunately, a commonly used log-log interpolation of tabulated EOSs can lead to significant numerical errors in the scalar derivatives of Eq.~(\ref{eq:scalar_charges}) because of its thermodynamic inconsistency.\footnote{Doing an interpolation of a tabulated EOS in a way that correctly accounts for the first law of thermodynamics is not a trivial problem and requires a formulation in terms of Helmholtz free energy \citep{swe96}. If one uses a simple linear interpolation in $\{\log n, \log p\}$ and $\{\log n, \log \varepsilon\}$ planes for a tabulated EOS ($n$ is the baryon number density and $\varepsilon$ is the energy density), the accuracy check can fail by the amount of $\delta_{\mathrm{rel}} \sim O(1) - O(0.1)$ in the region of strong scalarisation. The reason is that such interpolations do not obey the first law of thermodynamics $\frac{dn}{d\epsilon} = \frac{n}{\epsilon +p}$ which is assumed for structure equations and defines the baryonic mass $\Bar{m}_A$ \citep{har67,de96new}. As a consequence, this not only affects NS masses but even more so the derivatives in  Eq.~\eqref{eq:scalar_charges}, including the expression of $\alpha_A$ from Eq.~\eqref{eq:scalar_charges} as the derivative of the mass which enters the accuracy check.} For this reason, we use piece-wise polytropic approximations instead of tabulated EOSs, which is absolutely sufficient for conducting our tests and for exploring the EOS dependence of pulsar constraints on DEF gravity.

\section{A new timing model for scalar-tensor gravity}
\label{sec:DDSTG}

To place constraints on a gravity theory from the observations of a binary pulsar, one must know its predictions for the binary system's motion and the propagation of the electromagnetic signal in the curved spacetime of the binary system. The timing formula is the tool to capture the relativistic effects predicted by the theory from a sequence of pulse arrival times on Earth. The timing formula relates the observed (topocentric) time of arrival (TOA) of the pulse and its time of emission.

\subsection{Binary pulsar timing models}

The first timing model (BT) was introduced by \cite{bt76} in order to describe the timing of the first binary pulsar, PSR~B1913+16, which had been discovered earlier by \cite{ht75a}. The BT model assumes that the pulsar and its companion follow a Keplerian motion with additional secular changes in the Keplerian parameters of the orbit. utilised Keplerian parameters are the orbital period ($P_\mathrm{b}$), the epoch of periastron passage ($T_0$), the orbital eccentricity ($e$), the longitude of periastron ($\omega$), and the projected semi-major axis of the pulsar's orbit ($x$). The model accounts for a combination of a special-relativistic time dilation and a gravitational redshift; this periodic effect is described by an extra parameter called the Einstein delay ($\gamma$). The BT model accounts for (linear-in-time) secular changes in $P_\mathrm{b}$, $e$, $\omega$ and $x$, introducing PK parameters: rates of change of the orbital period ($\dot{P}_\mathrm{b}$), eccentricity ($\dot{e}$), longitude of periastron ($\dot{\omega}$) and projected semi-major axis ($\dot{x}$). These secular changes can be caused by both relativistic and astrometric effects \citep{dt92,lk04}.

Later, \cite{dd86} proved that all of the independent $O(v^2/c^2)$ timing effects could be described in a simple mathematical way for a wide range of alternative gravity theories. They developed a phenomenological (i.e.\ theory independent) timing model based on the full first post-Newtonian description of the two-body problem, which we will refer to as the ``DD'' model that uses a quasi-Keplerian solution to the equations of motion of a 2-body problem. This model allows working within a parametrised post-Keplerian approach (PPK). \cite{dt92} then showed that the DD model could be used to constrain a wide range of conservative theories of gravity obeying the modified Einstein-Infeld-Hoffmann (mEIH) framework (see also \citealt{wil93}). 

For the binary system part, the timing formula of the DD model reads as:
\begin{equation}
    \label{eq:timing_formula}
    t_\mathrm{b} - t_0 = D^{-1}\left[T + \Delta_R(T) + \Delta_E(T) + \Delta_S(T) + \Delta_A(T)\right] \, ,
\end{equation}
where $t_\mathrm{b}$ is the Solar System barycentric (infinite-frequency) arrival time, $t_0$ is a chosen reference epoch, $T$ is the pulsar's proper time, and $D$ is a Doppler factor accounting for the relative radial motion of the centre of mass of the binary system and the Solar System barycentre. The quantities $\Delta_i$ in Eq.~\eqref{eq:timing_formula} are different time delays introducing corrections due to internal binary effects.

Splitting the timing formula into a set of different contributions is to some extent a coordinate-dependent concept, however, within the approximations used it is convenient to work with these individual expressions for timing delays. The term $\Delta_R(T)$ is called ``Roemer delay'' and counts for the classical light travel time through the binary. $\Delta_E(T)$ is the ``Einstein delay'' and relates the proper emission time to the coordinate time of emission. $\Delta_S(T)$ is the ``Shapiro delay'' arising from the effect of the gravitational potential of the companion on the propagation of the pulsar signals. The ``aberration delay'' $\Delta_A(T)$ places corrections due to the periodic changes in the direction of pulse emission (as seen in the frame of the rotating pulsar) while the pulsar follows its binary motion.

All the expressions depend on three sets of parameters. Keplerian parameters are
\begin{equation}
    \label{eq:K_set}
    \{p^{\mathrm{K}}\} = \{P_\mathrm{b}, T_0, e_0, \omega_0, x_0\}
\end{equation}
and remain the same as for the BT model (the subscript 0 means a value at a given epoch). Separately measurable PK parameters are
\begin{equation}
    \label{eq:PK_set_sep}
    \{p^{\mathrm{PK}}\} = \{k, \gamma, \dot{P}_\mathrm{b}, r, s, \delta_{\theta}, \dot{e}, \dot{x}\} \, .
\end{equation}
The DD model introduces a periastron-shift parameter $k = \dot{\omega}P_\mathrm{b} / (2\pi )$, Shapiro ``shape'' ($s = \sin i$, where $i$ is the inclination angle) and Shapiro ``range'' ($r$) parameter for the signal propagation, and a relativistic deformation of the orbit ($\delta_{\theta}$). Not separately measurable PK parameters are
\begin{equation}
    \label{eq:PK_set_unsep}
    \{q^{\mathrm{PK}}\} = \{\delta_r, A, B, D\}\,,
\end{equation}
including a second parameter for the relativistic deformation ($\delta_{r}$), two aberration parameters ($A$ and $B$) and a Doppler factor ($D$).

The mentioned delays in the framework of the DD model are presented by expressions
\begin{align}
    \label{eq:delays}
    \Delta_R(T) &= x \sin \omega \left[\cos u - e (1 +\delta_r)\right]\nonumber\\
    & \quad\quad\quad + x \left[1 - e^2(1+\delta_\theta)^2\right]^{1/2} \cos \omega \sin u \,,\\
    \Delta_E(T) &= \gamma \sin u \,,\\
    \label{eq:delays_shapiro}
    \Delta_S(T) &= -2 r \ln \Big\{1 - e \cos u - s \big[\sin \omega (\cos u - e)\big.\Big.\nonumber\\
    &\quad\quad\quad \Big.\big. + (1 - e^2)^{1/2} \cos \omega \sin u \big]\Big\} \,,\\
    \Delta_A(T) &= A\left\{\sin\left[\omega + A_e(u)\right]+ e \sin \omega\right\}\nonumber\\
    \label{eq:delays_end}
    &\quad\quad\quad + B\left\{\cos\left[\omega + A_e(u)\right]+ e \cos \omega\right\} \,,
\end{align}
where the secular changes incorporated in the projected semi-major axis $x$ and eccentricity $e$ are given by
\begin{align}
    x &= x_0 + \dot x (T - T_0)\,,\\
    e &= e_0 + \dot e (T - T_0)\,.
\end{align}
The true anomaly ($A_e$) and $\omega$ depend on the eccentric anomaly ($u$) via following the relations
\begin{align}
    \label{eq:A_e}
    A_e(u) &= 2 \arctan \left[\left(\frac{1+e}{1-e}\right)^{1/2}\tan\frac{u}{2}\right] \,,\\
    \omega &= \omega_0 + k A_e(u) \,,
\end{align}
where $u$ is itself a function of $T$ and defined by solving Kepler's equation
\begin{equation}
    \label{eq:kepler}
    u - e \sin u = 2\pi \left[ \left(\frac{T-T_0}{P_\mathrm{b}}\right) - \frac12 \dot P_\mathrm{b} \left(\frac{T-T_0}{P_\mathrm{b}}\right)^2 \right]\,.
\end{equation}

The parameters in Eq.~\eqref{eq:PK_set_unsep} cannot be measured separately from the parameters in Eqs.~\eqref{eq:PK_set_sep} and \eqref{eq:K_set} as shown by \cite{dd86}. It means that all not separately measurable parameters can be fully absorbed into the change of other parameters. More details about measurable parameters and the connection between observed parameters and the intrinsic parameters can be found in \cite{dt92}. The redefinition of orbital parameters can absorb the Doppler factor ($D$). Such a redefinition does not affect the tests because it is only a rescaling of physical units, thus $D$ is set to 1.\footnote{Note, however, that a temporal change of $D$ is of relevance for pulsar timing experiments, as we will discuss later.}  The aberration parameters $A$ and $B$ can be absorbed as well by redefinition of $T_0, x, e, \delta_r$ and $\delta_{\theta}$.

\subsection{A timing model in DEF gravity}

The DD model can be applied to a wide range of fully-conservative theories of gravity. DEF gravity is a fully-conservative theory \citep{wil93}, thus we can investigate it in the framework of the DD model.

In different theories of gravity, the functional dependence of the PK parameters, i.e.
\begin{equation}
    \label{eq:PK_params}
    p^{\mathrm{PK}}_i = f_i^{\mathrm{theory}} (\massp, \mc; P_\mathrm{b}, e_0, x_0; \mathrm{EOS}) \,,
\end{equation}
can differ substantially because of strong-field effects involving a highly compact NS and its companion. The PK parameters are functions of two masses, thus one can perform gravity tests if at least 3 PK parameters are measurable. In general, if there are $n$ measured PK parameters, one can do $n-2$ independent tests for a given gravity theory. A common procedure of making tests using expressions for PK parameters is discussed in detail in Section~\ref{sec:J2222_now}, devoted to applications of a new model and comparison between different techniques.

The core part of the new timing model is the predictions for PK parameters (\ref{eq:PK_params}) in DEF gravity. We use expressions for PK parameters in DEF gravity provided in the literature \citep{de92a, de93new, de96new}. However, for parameters $\delta_{\theta}$, $\delta_r$, $A$ and $B$ there are only more general expressions. A phenomenological theory-independent description of PK parameters from which we derive their expressions in DEF gravity is presented in \citet{dt92}. The expressions of PK parameters and details are given in Appendix~\ref{App:PKs_in_DEF}. 

The aberration parameters $A$ and $B$ can also be calculated in DEF gravity, provided we know the system's geometry. The applications we use in the paper assume a special situation when the pulsar's spin axis is aligned to the orbital angular momentum of the system. The alignment is a reasonable assumption for a recycled pulsar with a WD companion due to the preceding accretion process during the system's evolution. In this special case, $A$ is calculated according to Eq.~\eqref{eq:ppk_a} assuming angles $\eta = -\pi/2$ and $\lambda = i$ \citep{dt92}. The second aberration parameter $B = 0$ in this special situation. In general, an alignment may not be the case for a particular binary pulsar (e.g. double neutron star systems), but the new model can also handle this misalignment. Once the system's geometry is assumed, real $A$ and $B$ values can be calculated for a given epoch without the necessity to redefine other parameters.\footnote{If the spin of the pulsar is misaligned, this can lead to a further complication of the analysis. In such a case geodetic precession leads to apparent changes in the binary parameters \citep{dt92,lk04}.}

The PK parameters from Eq.~\eqref{eq:PK_params} depend on the orbital parameters of the system and the physical properties of the pulsar. However, we also have to know properties of the companion. The PK expressions (\ref{eq:ppk_gamma} - \ref{eq:ppk_pbdot_quadg}) depend on gravitational form-factors of the companion $\{\alphac, \betac\}$. The quantity $k_\mathrm{c}$ does not appear in the expressions because they take into account only leading order terms, and is therefore of no interest. These quantities, in turn, depend on the type of companion. If the companion is a NS, they are calculated in the same way as for the primary pulsar using Eq.~\eqref{eq:scalar_charges}. If the companion is a black hole (BH), then the gravitational form-factors all vanish:
\begin{equation}
    \alpha_{\mathrm{BH}} = 0,\quad \beta_{\mathrm{BH}} = 0,\quad k_{\mathrm{BH}} = 0\,.
\end{equation}
This is a consequence of the ``no-hair'' theorem; the BH is not scalarised in DEF gravity \citep{haw72,de92}. Section~\ref{sec:PSRBH} is dedicated to a more thorough discussion on binary pulsar systems with a BH and the results which we can obtain from timing of a pulsar in a relativistic orbit with a BH. For a WD companion, the gravitational form-factors are approximated by their weak field expressions
\begin{equation}
    \alpha_{\mathrm{WD}} \simeq \alpha_0,\quad \beta_{\mathrm{WD}} \simeq \beta_0,\quad k_{\mathrm{WD}} \simeq 0\,.
\end{equation}
Such an approximation is sufficient for our purposes because WDs are not compact enough to show the strong field effects present in NSs and BHs. The next order approximation for weakly self-gravitating objects is 
$\alpha_{A} \simeq \alpha_0 (1 - 2 s_A)$, 
where $s_A \simeq G_* m_A/(Rc^2)$ is the sensitivity ($R$ is the object's radius) and has a typical value of $s_{\mathrm{WD}} \sim 10^{-5} - 10^{-3}$ for a WD \citep{de92,de93new,wil18}. Thus the usage of weak field counterparts $\{\alpha_0, \beta_0\}$ for a WD companion instead of precisely calculated values is justified.

To summarise, the timing model in DEF gravity is defined by two theory parameters $\{\alpha_0, \beta_0\}$, the chosen EOS for a NS, and the type of the companion among \{WD, NS, BH\}. We name the new model DDSTG arises as direct extension of the DD model for STG theories.

\subsection{DDSTG implementation into \textsc{TEMPO}}

Once we obtain the theoretical part of the new DDSTG timing model, we need to apply it to the timing data. We implemented the DDSTG model into one of the commonly used pulsar timing software -- the \textsc{TEMPO}\footnote{\label{url_tempo}\url{http://tempo.sourceforge.net}} program \citep{nds15}. The local implementation of DDSTG model in \textsc{TEMPO} and precalculated grids of masses and gravitational form-factors are supplied with the paper\footnote{\label{url_ddstg}\url{https://github.com/AlexBatrakov/tempo-13.103_ddstg}}. The authors intend to make DDSTG a part of the official \textsc{TEMPO} distribution to ensure forward compatibility.

There is a standard procedure of analysing radio pulsar timing data. The preprocessed TOAs are obtained after radio observations of the desired pulsar system. During the procedure, \textsc{TEMPO} reads TOAs, parameters of the binary model, and some coded instructions from supplied files. Then \textsc{TEMPO} fits the selected timing model accounting for the transformation to the Solar System barycentre, pulsar rotation, and its spin down for a chosen binary model.

Specifically for DDSTG, a user selects theory parameters $\{\alpha_0, \beta_0\}$, EOS, and the type of a companion. During the initialisation \textsc{TEMPO} reads pre-calculated 3D grids of gravitational form-factors and NS masses $\{\alpha_A, \beta_A, k_A, m_A\}$ (see Section \ref{subsec:calc_grids}) for the chosen EOS. Each 3D grid depends on $\{\alpha_0,\beta_0,p_c\}$ parameters, where $p_c$ is the central pressure. Then gravitational form-factors and masses are interpolated for the selected $\{\alpha_0, \beta_0\}$ values and saved into a smaller 1D grids (depending only on $p_c$) in the program memory. The final 1D grids remain unchanged and are used to interpolate gravitational form-factors for a particular theory in the mass range during the fitting procedure of \textsc{TEMPO}. 

\textsc{TEMPO} with the selected DDSTG model fits two masses: the total mass ($m_{\mathrm{tot}}$) of the system and the companion's mass ($m_{\mathrm{c}}$). 
Every time when these masses change, the model recalculates the gravitational form-factors for the pulsar and the companion. Once the gravitational form-factors (Eq.~\eqref{eq:scalar_charges}) are known, the model calculates all PK parameters (Eqs.~\eqref{eq:PK_set_sep} and \eqref{eq:PK_set_unsep}) using equations given in Appendix~\ref{App:PKs_in_DEF}. These calculated PK parameters are used afterwards in the timing formula \eqref{eq:timing_formula}. The details about the DDSTG implementation and the description of model parameters can be found in Appendix \ref{App:DDSTG_implemetation}.

In the timing formula \eqref{eq:timing_formula}, the Solar System barycentric arrival time ($t_\mathrm{b}$) is known. However, to obtain the proper pulsar time ($T$), one has to perform the inversion of the timing formula, i.e.\ get $T$ as a function of $t_\mathrm{b}$: $T = t_\mathrm{b} - \bar\Delta(t_\mathrm{b})$. The original DD model utilises an approximate analytic inversion, which sometimes is not accurate enough, for example, for the Double Pulsar \citep{ksm21}. In contrast, our DDSTG model performs accurate numerical inversion. Eq.~\eqref{eq:timing_formula} is solved iteratively for $T$ for each TOA. The discrepancy due to an inaccurate inversion may influence the test for precise timing in the future or PSR-BH systems discussed further below. Nowadays, numerical inversion is considered a new standard and implemented in the DDS model of \textsc{TEMPO} \citep{ksm21} and generally used in PINT \citep{pint21}.

The DDSTG model produces the best fit to the data for DEF gravity with a selected set of parameters and a given EOS. The output of \textsc{TEMPO} stays the same as for other binary models, e.g. the calculated $\chi^2$ value and root mean square error of fit. The $\chi^2$ statistics may be applied for further tests to compare the results for different theory parameters.

\subsection{Advantages of DDSTG}\label{subsec:advantages_ddstg}

\begin{figure*}
   \centering
   \includegraphics[width=9cm]{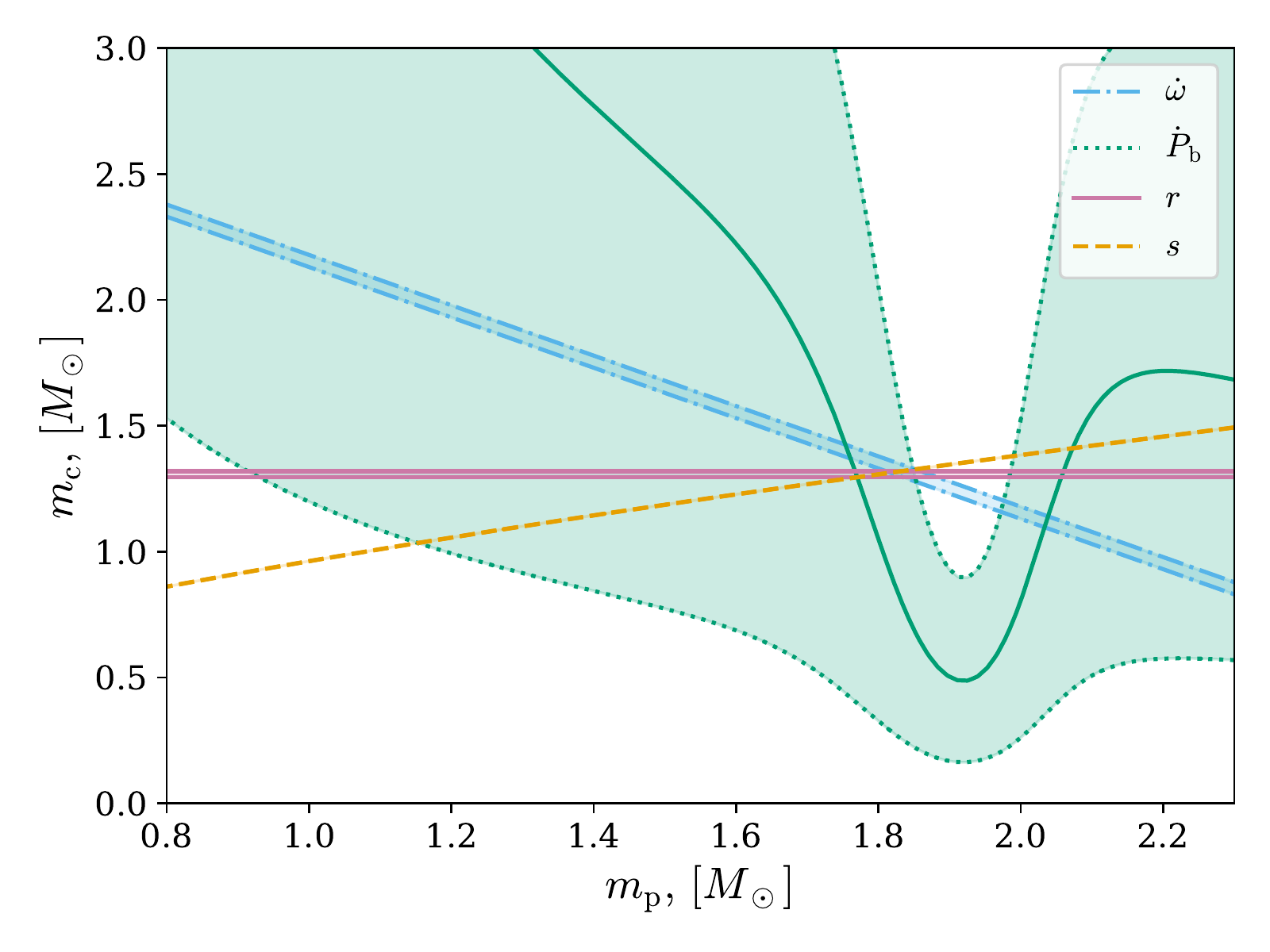}
   \includegraphics[width=9cm]{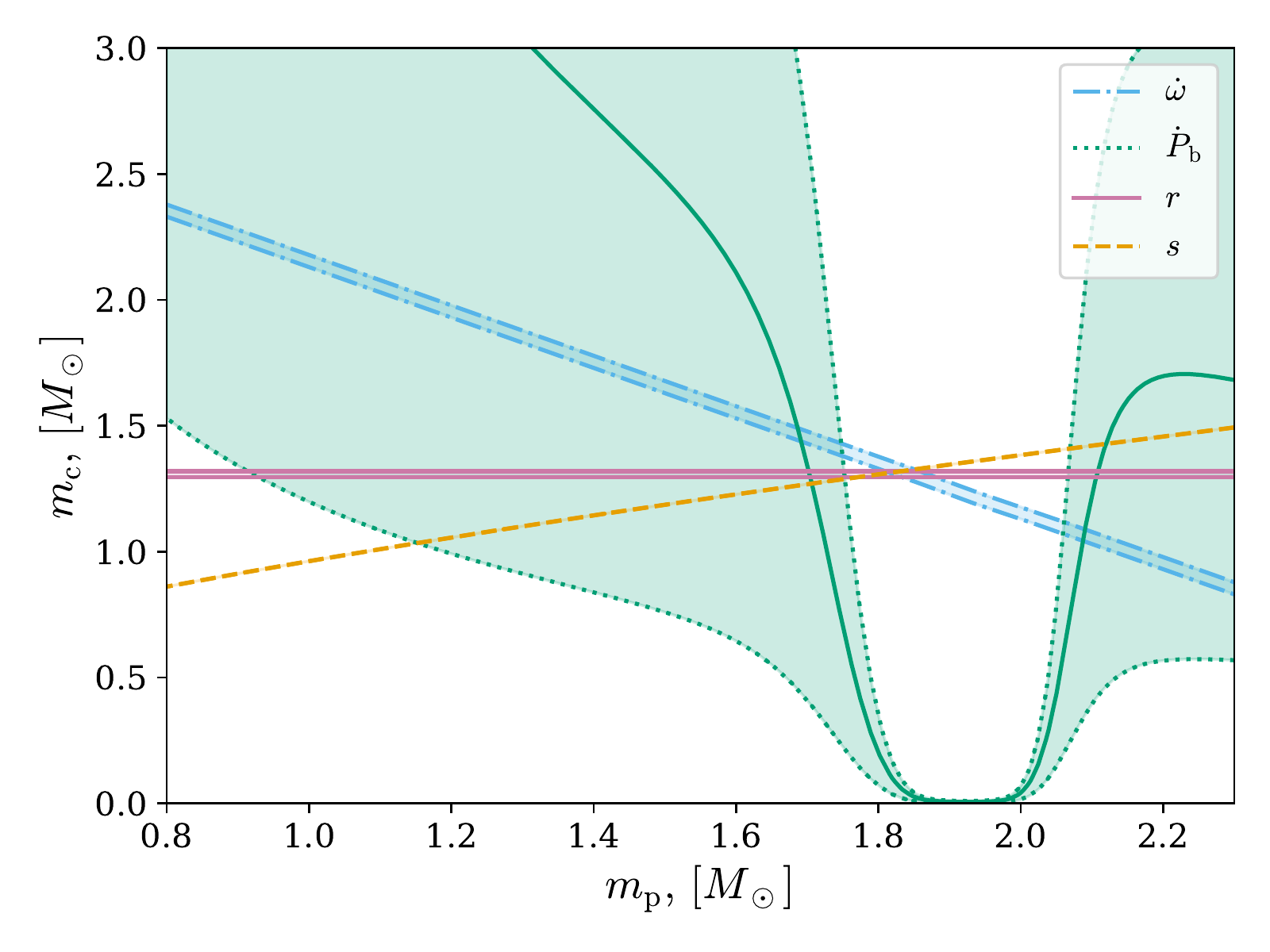}
   \caption{Mass-mass diagrams for the PSR~J2222$-$0137 system at the zone of high nonlinearity where a fast transition to a strongly scalarised pulsar happens. PK parameters are obtained by fitting the DD model to the timing data from \cite{gfg21}. Calculations are preformed within DEF gravity for the MPA1 EOS allowing NSs with maximum mass in GR of $M_\mathrm{GR} \simeq 2.46 M_\odot$. Left panel corresponds to the point in the DEF parameter space with $\alpha_0 = -10^{-4}, \beta_0 = -4.3$, which is not excluded by the test. Right panel shows the same test but for $\alpha_0 = -10^{-4}, \beta_0 = -4.35$. The prediction of a strong dipolar contribution $\dot{P}_\mathrm{b}^D$ fails the test for the more negative $\beta_0 = -4.35$. The axes correspond to the masses of the pulsar $m_\mathrm{p}$ and the companion $\mathrm{c}$. The shadowed area is the allowed region at $68\%$ CL limit for a corresponding PK parameter. The solid green line corresponds to the observed value of $\dot{P}_\mathrm{b}^{GW}$ due to the GW emission.}
    \label{Fig:J2222_MM_DEF_scalarisation}
   \end{figure*}

The tests with pulsar data allow placing constraints on the gravity theory parameters. The most simple and common way to achieve this is to compare the measured PK parameters with the theoretical predictions from the theory. PK parameters are obtained by fitting a phenomenological model to the observational data, e.g., DD model. A phenomenological model estimates PK parameters and their uncertainties.

Within a particular gravity theory, each PK parameter depends on the masses of the pulsar and the companion and thus corresponds to a curve in a mass-mass space. Together with the measurement error, each PK parameter produces a strip in the $\{\massp, \mc\}$ space. If there are two measured parameters, one may find the intersection area and obtain the estimated mass values. If the number of measured parameters is more than two, one can perform a test of that particular gravity theory. The test is passed if all three curves intersect in the range of errors at one point. Generally, $n$ measured PK parameters result in $n-2$ independent tests.

If a gravity theory has arbitrary parameters, tests can be done for any fixed values of theory parameters. For DEF gravity, this procedure results in an ``allowed'' area in $\{\alpha_0, \beta_0\}$ space that pass gravity tests within the measurement uncertainties. This area is bounded within the selected confidence limit by a curve, which is usually plotted. To date, the regions allowed by different tests have always included GR.

If the companion is optically bright, one can get information about the system from optical observations. For example, for a binary pulsar system with a bright WD it is possible to obtain the mass ratio and the companion mass using high-resolution optical spectroscopy of the companion (e.g., \citealt{avk+12,afw+13}). In such cases, we do not have to measure three independent PK parameters based on the radio data, but we can combine several multi-wavelength constraints.

Another issue are possible correlations between PK and other parameters. Observed correlations can come from the theoretical correlation of parameters within the binary model (e.g. $T_0 \leftrightarrow \omega$ and $P_\mathrm{b} \leftrightarrow \dot{\omega}$ in a low eccentric case) and from a nonuniform data sampling (e.g. a nonlinear correlation in a parametrisation of Shapiro delay $r \leftrightarrow s$). Often in the past, measured PK parameters were published without any information about observed correlations. This additional information from the timing data is lost in such cases. These days it became a common practice to publish observed correlations and even provide them explicitly. \cite{and19} and \cite{afy19} accounted for possible observed and theoretical correlations between PK parameters within DEF and MO gravity via computationally highly demanding Markov chain Monte Carlo (MCMC) simulations utilising the correlation matrix from the \textsc{TEMPO} output. However, this elaborate approach would not work if the relativistic effect is not well measured or the dependence between the parameters is highly nonlinear. If the relation is nonlinear, the correlation matrix gives information only about the linear contribution. To fully account for a nonlinear relation one has to obtain a full probability density function in the parameter space.

Contrary to the PK method, the DDSTG model accounts for all even nonlinear correlations naturally and breaks them by a direct fit of the masses. This was already one of the main advantages of the DDGR model; and is one of the reasons why the DDSTG uses the same superior approach while extending it to STG theories. The model accounts for all effects internally and extracts all the information from the timing data. The information is not lost even if the correlating parameter cannot be measured but influences other parameters. Such possible correlations are relevant for PSR~J1141$-$6545 \citep{vvk19,vbv+20}, where there is a correlation between the time dilation parameter ($\gamma$) of the Einstein delay  and the rate of change of the projected semi-major axis due to the spin-orbit coupling caused by the fast rotating WD companion ($\dot{x}^{\mathrm{SO}}$). Furthermore, the Shapiro delay cannot be measured independently but is still essential \citep[see e.g.][]{bbv08}. We expect the DDSTG model to be of particular advantage for PSR~J1141$-$6545 and there will be a  dedicated paper (Venkatraman Krishnan et al. in prep.) about applying the new model to this system.

Compared to conventional methods, the constraints obtained by DDSTG may, in general, become either tighter or weaker depending on the particular case. However, the resulting restrictions are more reliable by the construction. This weakening of the limits can happen because of unaccounted correlations.

\section{Application to PSR~J2222\texorpdfstring{$-$}{--}0137}
\label{sec:J2222_now}

As a demonstration of the DDSTG model, we now apply it to published timing data of a binary pulsar. We select PSR~J2222$-$0137 because of its unusual characteristics, which we now list in detail.

\subsection{About the system}

We are particularly interested in recycled pulsars because of their, in general, better timing precision essential for precise gravity tests. The radio pulsar PSR~J2222$-$0137 was discovered in the Green Bank Telescope (GBT) $350$ MHz drift scan pulsar survey \citep{blr+13}. It is a mildly recycled pulsar which has a spin period ($P$) of $32.8$ ms. The pulsar is in a binary system with $P_\mathrm{b}$ of $2.44576$ days and $x$ of $10.848$ light-seconds. We also expect the pulsar's spin axis to be aligned to the orbital angular momentum of the system. The alignment happens as a consequence of the pulsar recycling process, when the pulsar accretes matter from the companion.

PSR~J2222$-$0137 is already known as a unique laboratory for testing gravity theories because of its special characteristics; for more detailed information, we refer the reader to \cite{cfg+17} and \cite{gfg21}. It is one of the closest pulsars known, with the most precise distance measured with Very Large Baseline Interferometry (VLBI, see \citealt{dbl+13}) and excellent timing precision. The system has a highly significant detection of the Shapiro delay as well as the measured rate of advance of periastron ($\dot{\omega}$), which yield precise mass measurements and $\sim 1$\% test of the GR predictions for the Shapiro delay \citep{gfg21}.

The most important characteristic in the scope of this paper is the precise measurement of $\dot{P}_\mathrm{b}$. Given the precise masses, this can be compared with a precise prediction for the orbital decay due to GW damping, furthermore the kinematic contributions to $\dot{P}_\mathrm{b}$ (see Section~\ref{subsec:PBDOT_uncertainties}) can be estimated precisely because of the good distance measurement. The $\dot{P}_\mathrm{b}$ measurement constrains dipolar GW emission in this system, which would arise in DEF gravity \citep{de92} because of the large difference in the compactness of the components (NS and WD). Finally, the mass of the pulsar ($m_{\mathrm{p}} = 1.831(10) M_{\odot}$) \citep{gfg21} lies in the range where spontaneous scalarisation happens, yielding strong limits on this highly non-linear phenomenon \citep{ssb+17,zfk+22}.

\subsection{Observational data}
\label{sec:J2222_obs}

The PSR~J2222$-$0137 timing data used in this work are the same as used by \cite{gfg21}. This includes observations of this pulsar going back to its original follow-up, which started on 2009 June 23 using the Green Bank Telescope (GBT); however these ended on 2011 December 26. Regular observations with the three largest European radio telescopes (3ERT) started the following years: the Nan\c{c}ay Radio Telescope on 2012 October 4, The Lovell Telescope at Jodrell Bank on 2012 November 20 and the dense orbital campaigns with the Effelsberg 100-m radio telescope on 2015 October 26. These observations continue to the present day; however \cite{gfg21} only analysed the data obtained until 2021 May 2; their processing and how the resulting ToAs were analysed is described in detail by \cite{cfg+17} and \cite{gfg21}.

Observations with the MeerKAT radio telescope array and the Five hundred meter Aperture Spherical Telescope (FAST) started in 2019 September 24 and 2020 October 5, but these timing data were not used by \cite{gfg21}. However, the latter authors did use FAST data for a detailed study of the emission properties of PSR~J2222$-$0137. In this work, we also extend the existing data by simulated data assuming the same timing properties of all of these telescopes to simulate how the timing parameters for this system will improve in the near future (see Section~\ref{sec:J2222_future}).

\subsection{Mass-mass diagrams in DEF gravity}

The illustrative and straightforward way to explore if a given gravity theory agrees with binary pulsar observational data is to calculate mass-mass diagrams. As mentioned in Section~\ref{subsec:advantages_ddstg}, firstly, one measures values of PK parameters with their uncertainties by fitting the phenomenological binary model DD to binary pulsar data. Then the observed PK values can be compared with their theoretical predictions of a specific gravity theory. The PK parameters from Eq.~\eqref{eq:PK_params} depend on the mass of the pulsar ($m_{\mathrm{p}}$) and the mass of the companion ($m_{\mathrm{c}}$) (see Appendix~\ref{App:PKs_in_DEF}). For PSR~J2222$-$0137 we use the measured PK parameters from \cite{gfg21}.

To illustrate how the tests are performed, we select two points in DEF gravity parameter space near the scalarisation region with large negative $\beta_0$ and calculate mass-mass diagrams for them. The test is sensitive to the predicted dipolar contribution to $\dot{P}_\mathrm{b}$, which rises dramatically in the  region of spontaneous scalarisation. The first point with $\alpha_0 = -10^{-4}, \beta_0 = -4.3$ does not predict a strong enough dipolar contribution and passes the test within a $1\sigma$ limit. The corresponding mass-mass diagram is presented in the left panel of Figure~\ref{Fig:J2222_MM_DEF_scalarisation}. In contrast, the second point with $\alpha_0 = -10^{-4}$ and a bit smaller $\beta_0 = -4.35$ is already excluded because of a rather strong scalarisation of the pulsar leading to significant dipolar GW damping (see the right panel of Figure~\ref{Fig:J2222_MM_DEF_scalarisation}). The $\dot{P}_\mathrm{b}$ curve nicely shows that a significant enhancement in the scalarisation happens for a particular interval in the mass range defined by the EOS.

\subsection{Various contributions to \texorpdfstring{$\dot{P}_\mathrm{b}$}{PBDOT}}
\label{subsec:PBDOT_uncertainties}

At this point we need to discuss the different contributions to the observed $\dot{P}_\mathrm{b}^{\mathrm{obs}}$ and their influence on our results. The observed orbital decay of the system consists of many terms
\begin{equation}\label{eq:pb_obs}
    \dot{P}_\mathrm{b}^{\mathrm{obs}} = \dot{P}_\mathrm{b}^{\mathrm{GW}} + \dot{P}_\mathrm{b}^{\mathrm{Gal}} + \dot{P}_\mathrm{b}^{\mathrm{Shk}} + \dot{P}_\mathrm{b}^{\mathrm{Tid}} + \dot{P}_\mathrm{b}^{\dot{M}} + \dot{P}_\mathrm{b}^{\dot{G}} \, ,
\end{equation}
where the first term $\dot{P}_\mathrm{b}^{\mathrm{GW}}$ is due to GW damping and can include dipolar and monopolar terms in DEF gravity, besides the general quadrupolar prediction of GR. The full expressions of different GW contributions are presented in Appendix~\eqref{eq:ppk_pbdot}. The next two terms are of kinematic origin: the Shklovskii effect ($\dot{P}_\mathrm{b}^{\mathrm{Shk}}$) and the Galactic contribution ($\dot{P}_\mathrm{b}^{\mathrm{Gal}}$). They are the result of a time-varying Doppler factor $D$ due to an (apparent) radial acceleration between the pulsar binary and the Solar System \citep{dt91}. The last three terms come from tidal effects, mass loss in the system, and a possible temporal variation of the gravitational constant $G$.

We are mainly interested in measuring the GW emission term ($\dot{P}_\mathrm{b}^{\mathrm{GW}}$) and comparing it with the prediction of DEF gravity. The most significant additional effects in PSR~J2222$-$0137 come from kinematic contributions $\dot{P}_\mathrm{b}^{\mathrm{Gal}}$ and $\dot{P}_\mathrm{b}^{\mathrm{Shk}}$. These two effects arise beyond the binary system and can be combined in the overall external contribution
\begin{equation}\label{eq:Pbdot_ext}
    \dot{P}_\mathrm{b}^{\mathrm{ext}} = \dot{P}_\mathrm{b}^{\mathrm{Gal}} + \dot{P}_\mathrm{b}^{\mathrm{Shk}} \,.
\end{equation}
The last three terms in Eq.~\eqref{eq:pb_obs} are parts of the internal contribution
\begin{equation}
    \dot{P}_\mathrm{b}^{\mathrm{int}} = 
    \dot{P}_\mathrm{b}^{\mathrm{GW}} +
    \dot{P}_\mathrm{b}^{\mathrm{Tid}} + \dot{P}_\mathrm{b}^{\mathrm{\dot{M}}} + \dot{P}_\mathrm{b}^{\mathrm{\dot{G}}} \,,
\end{equation}
and can be neglected for PSR~J2222$-$0137. Thus the internal contribution has only one significant term left, the GW term. For the discussion about possible account of $\dot{P}_\mathrm{b}^{\dot{G}}$ we refer the reader to Section~\ref{sec:gdot}. Within the DDSTG model, the parameter XPBDOT in \textsc{TEMPO} is used for treating both external and internal effects $\dot{P}_\mathrm{b}^{\mathrm{ext}} + \dot{P}_\mathrm{b}^{\mathrm{int}} - \dot{P}_\mathrm{b}^{\mathrm{GW}}$ simultaneously subtracting the GW term. In our case XPBDOT accounts for only external Shklovskii and Galactic effects.

There is an extensive analysis of the external terms for PSR~J2222$-$0137 by \cite{gfg21} and their determined values are: $\dot{P}_\mathrm{b}^{\mathrm{Gal}} = -0.0142(13)\times 10^{-12}$ s\,s$^{-1}$ and $\dot{P}_\mathrm{b}^{\mathrm{Shk}} = 0.2794(12)\times 10^{-12}$ s\,s$^{-1}$. These values correspond to the external variation of the observed orbital period of
\begin{equation}
    \dot{P}_\mathrm{b}^{\mathrm{ext}} = 0.2652(18)\times 10^{-12}\, \mathrm{s\,s}^{-1} \, ,
\end{equation}
which is consistent within $2\sigma$ with the total observed value
\begin{equation}
    \dot{P}_\mathrm{b}^{\mathrm{obs}} = 0.2509(76)\times 10^{-12}\, \mathrm{s\,s}^{-1} \, ,
\end{equation}
leaving the internal contribution $\dot{P}_\mathrm{b}^{\mathrm{int}} = \dot{P}_\mathrm{b}^{\mathrm{obs}} - \dot{P}_\mathrm{b}^{\mathrm{ext}} = -0.0143(78)\times 10^{-12}\, \mathrm{s\,s}^{-1}$ consistent with the GR prediction for quadrupolar GW emission $\dot{P}_\mathrm{b}^{\mathrm{GR}} = -0.00809(5)\times 10^{-12}\, \mathrm{s\,s}^{-1}$ \citep{gfg21}.

According to \cite{dt91}, \cite{nt95}, and \cite{lwj+09}, the Galactic differential acceleration may be analytically approximated with the expression
\begin{equation}\label{eq:pbdot_gal}
    \frac{\dot{P}_\mathrm{b}^{\mathrm{Gal}}}{P_\mathrm{b}} = - \frac{K_z |\sin{b}|}{c} - \frac{\Theta^2_0}{c R_0}\left(\cos{l}+\frac{\beta}{\beta^2 + \sin^2{l}}\right)\cos{b} \, ,
\end{equation}
where $l$ is Galactic longitude, $b$ the Galactic latitude and $\beta = (d/R_0)\cos{b} - \cos{l}$. The quantity $K_z$ is the vertical component of the Galactic acceleration, which for Galactic heights $z \equiv |d \sin b| \leq 1.5$ kpc can be approximated with sufficient accuracy by
\begin{equation}
    K_z(10^{-9}\ \mathrm{cm\,s}^{-2})\simeq 2.27 z_{\mathrm{kpc}}+3.68(1-e^{-4.31z_{\mathrm{kpc}}}) \, ,
\end{equation}
where $z_{\mathrm{kpc}} \equiv z(\mathrm{kpc})$ \citep{hofl04,lwj+09}. From \cite{Gravity2021} we can take a value for the Sun’s Galactocentric distance $R_0 = 8275 \pm9 \pm33$ pc. The Galactic circular velocity at the location of the Sun ($\Theta_0$) is taken to be $240.5(41)$ km/s (see \citealt{gfg21}, and references therein).

The Shklovskii contribution \citep{shk70} can be calculated using
\begin{equation}\label{eq:pbdot_shk}
    \dot{P}_\mathrm{b}^{\mathrm{Shk}} = \frac{(\mu^2_{\alpha} + \mu^2_{\delta})\,d}{c} \, P_\mathrm{b} \,,
\end{equation}
where $\mu_{\alpha}$ and $\mu_{\delta}$ are the proper motion in Right Ascension (RA) and Declination, respectively, and $d$ is the distance to the pulsar. The astrometric values and uncertainties for PSR~J2222$-$0137 are taken from \cite{gfg21}. For PSR~J2222$-$0137, the uncertainty in $\dot{P}_\mathrm{b}^{\mathrm{ext}}$ is small compared to the precision of the $\dot{P}_\mathrm{b}$ measurement. The uncertainty of $\dot{P}_\mathrm{b}^{\mathrm{ext}}$ in our case can be neglected and therefore we can provide the fixed XPBDOT parameter in \textsc{TEMPO}.

\subsection{Results of applying DDSTG}

   \begin{figure}
   \centering
   \includegraphics[width=9cm]{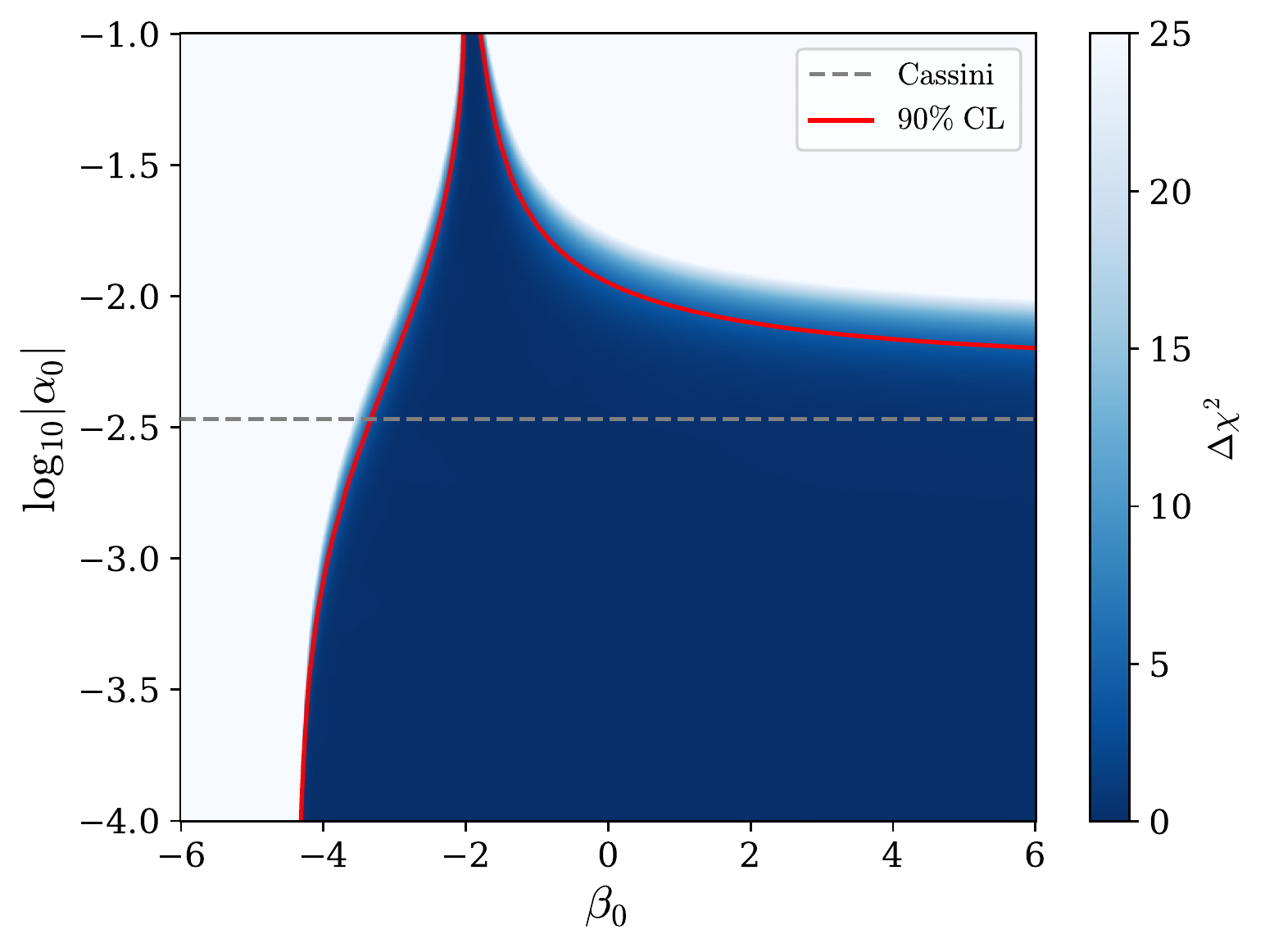}
   \caption{$\Delta\chi^2$ map in DEF gravity parameter space for PSR~J2222$-$0137. The test is performed by applying the DDSTG model to the timing data of \cite{gfg21} and the assuming rather stiff MPA1 EOS. The red line corresponds to $90\%$ CL limit ($\Delta\chi^2 \simeq 4.6$), the area above the grey line is restricted by Cassini mission. GR with $\alpha_0 = 0, \beta_0 = 0$ lies beyond the plotted domain in the blue region at the bottom at the infinity.}
    \label{Fig:J2222_chisqr}
    \end{figure}
%

   \begin{figure}
   \centering
   \includegraphics[width=9cm]{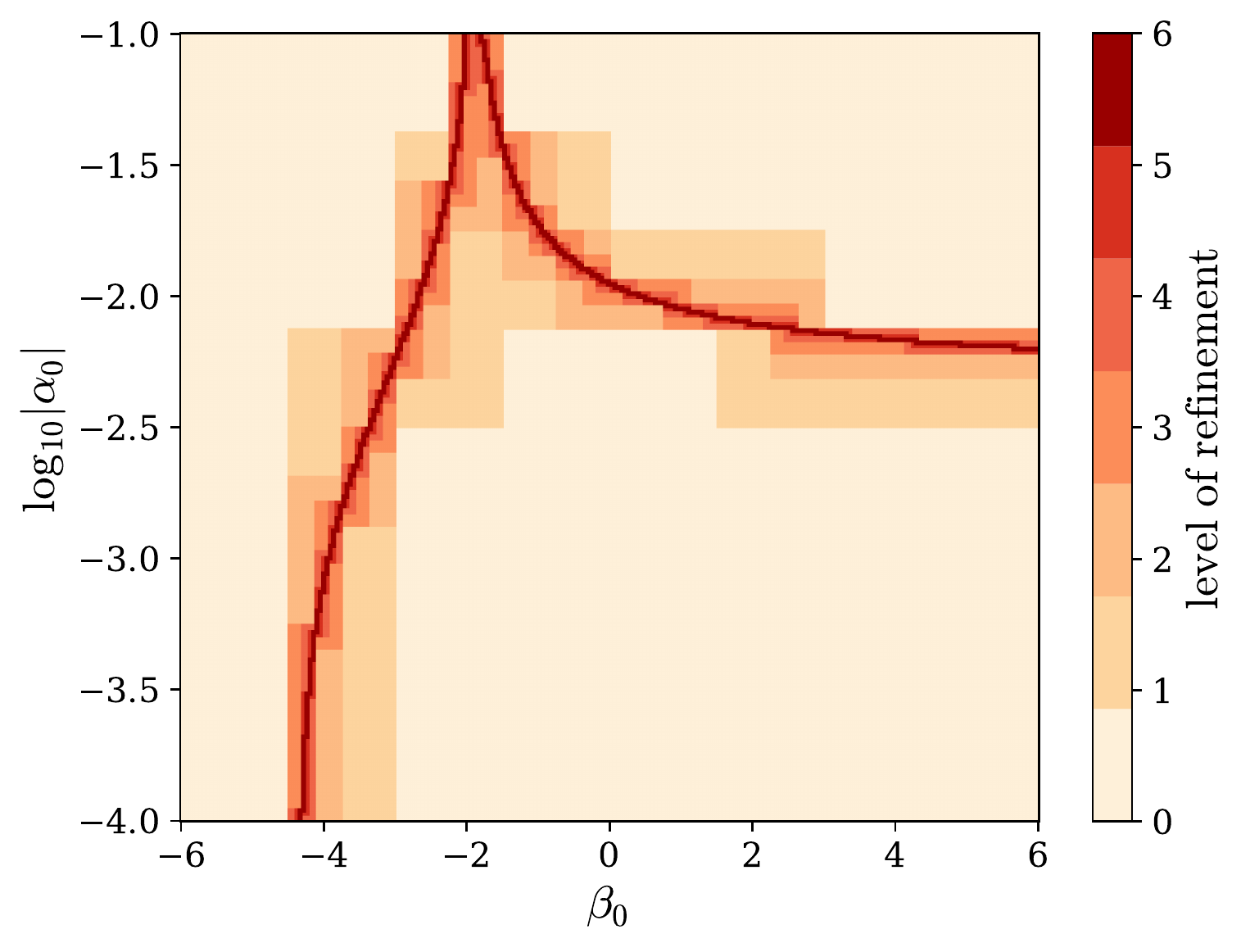}
      \caption{Map of the refinement level in DEF parameter space. The test is the same as in Figure~\ref{Fig:J2222_chisqr}. Areas in the parameter space with a higher refinement level have an exponentially growing resolution. The adaptive refinement procedure resolves the contour of $90\%$ CL limit and saves computational time dramatically.
              }
         \label{Fig:J2222_ref_level}
   \end{figure}

Our main goal is to obtain limits on DEF gravity parameters $\{\alpha_0, \beta_0\}$ by applying the DDSTG model. \textsc{TEMPO} allows to calculate a $\chi^2$ value for every particularly selected pair of values $(\alpha_0, \beta_0)$. Therefore we obtain $\chi^2$ values for a grid of parameters and compare them to each other. For straightforward comparison we subtract the minimum $\chi^2_{\mathrm{min}}$ over the $\{\alpha_0, \beta_0\}$ grid from calculated $\chi^2$. The location of the minimum $\chi^2_{\mathrm{min}}$ is statistically in agreement with GR $(\alpha_0=0, \beta_0 = 0)$. The shifted quantity $\Delta\chi^2 = \chi^2 -\chi^2_{\mathrm{min}}$ has a $\chi^2(\mathrm{d.o.f.} = 2)$ distribution with 2 degrees of freedom. Finally we construct contours of a fixed $\Delta\chi^2$ value corresponding to desired confidence levels. In this paper we adhere to $90\%$ confidence level limits with $\Delta\chi^2 \simeq 4.6$. 

During each run, \textsc{TEMPO} fits spin parameters, astrometric parameters, Keplerian parameters, two masses, and other parameters of interest, e.g. $\dot{x}$. A value of the external contribution to the rate of the orbital period change ($\dot{P}_\mathrm{b}^{\mathrm{ext}}$) is fixed and selected, so that it fully accounts for Shklovskii effect and the Galactic $\dot{P}_\mathrm{b}$ contributions \citep{shk70,bzn+75,dt92}.  The resulting $\Delta\chi^2$ map in $\{\alpha_0, \beta_0\}$ space for PSR~J2222$-$0137 is presented in Figure~\ref{Fig:J2222_chisqr}. The limit on DEF parameters is placed by a contour of $90\%$ confidence level limit of $\chi^2(\mathrm{d.o.f.} = 2)$ statistics.

Generally, the uncertainty of $\dot{P}_\mathrm{b}^{\mathrm{ext}}$ can be important for other systems where it is not well constrained. In this case, the DDSTG model allows a complete description with account for all the uncertainty due to external effects $\dot{P}_\mathrm{b}^{\mathrm{ext}}$. First we have to assume a prior distribution for the $\dot{P}_\mathrm{b}^{\mathrm{ext}}$ value. Then we have to calculate the $\chi^2$ value for a 3-dimensional map with $\{\alpha_0, \beta_0, \dot{P}_\mathrm{b}^{\mathrm{ext}}\}$ as parameters of the three axes. Finally, we marginalise the probability density for the $\dot{P}_\mathrm{b}^{\mathrm{ext}}$ axis using Bayes' theorem and obtain the corrected 2-dimensional map of $\chi^2_{\mathrm{corr}}$. The details of the procedure are presented in Appendix~\ref{App:XPBDOT}.

Calculating an extensive grid in a 2-dimensional parameter space with a lot of points (e.g. $\sim 500$) for each axis to obtain a finely resolved contour of the desired $\Delta\chi^2$ value is too computationally demanding. In this work, we use an adaptive mesh refinement technique to trace the location of the desired contour. We start from the sparse grid with $9\times9$ points in the $\{\alpha_0, \beta_0\}$ space. Then, the algorithm resolves all the cells that can have the contour line inside them. The algorithm repeats the refinement of important cells until we obtain the desired curve fineness. 

For a $N\times N$ grid, the naive approach utilises $O(N^2)$ iterations, whereas adaptive mesh refinement has only $O(N \log_2 N)$ complexity. For the present work we use the refinement level of 6, resulting in the final grid with $513\times513$ points. We need such a large resolution in the parameter space because of the ``horn'' feature at large $\alpha_0$ values near $\beta_0 \simeq -2$. A simple analysis shows that adaptive refinement requires to calculate 56 times fewer points to resolve a single contour for a grid with $513\times513$ points. In Figure~\ref{Fig:J2222_ref_level} we present an adaptive refinement map corresponding to the search of a contour from Figure~\ref{Fig:J2222_chisqr}. A high level of refinement is performed only along the contour of $90\%$ CL limit. In Appendix~\ref{App:MM_horn} one can also find the mass-mass diagram showing what happens in the region of the ``horn'' in terms of PK parameters.

\subsection{Comparison with the PK method}

In this section we compare the constraining power of the newly developed approach of this paper with that of existing procedures. For this reason we perform the test with the traditional ``PK method'' based on the PK parameters from the timing data. The corresponding PK parameters are measured by means of fitting the DD model \citep{dd86} to the same timing data. Then for each specific choice of the theory parameters $\{\alpha_0, \beta_0\}$ we fit two masses $\massp$ and $\mc$ to minimise the $\chi^2$ value. The $\chi^2$ is calculated by comparison of the observed PK parameters $p^\mathrm{obs}_i$ and predicted values $p^\mathrm{theory}_i$ from the theory
\begin{equation}
    \chi^2(p^{\mathrm{obs}}_i; \massp,\mc) = \sum_i \left(\sigma^\mathrm{obs}_{p_i}\right)^{-2}\left(p^\mathrm{theory}_i(\massp,\mc) - p^{\mathrm{obs}}_i\right)^2 \, ,
\end{equation}
where $\sigma^\mathrm{obs}_{p_i}$ is the standard deviation of the i-th measured PK parameter. Finally, we calculate the grid of $\chi^2$ values over the desired $\{\alpha_0, \beta_0\}$ space with the same settings as for the DDSTG approach and shift $\chi^2$ values by its minimum $\chi^2_\mathrm{min}$. The detailed explanation of this common ``PK method'' can be found in \citet{de98}. For both methods the minimum $\chi^2_\mathrm{min}$ is in statistical agreement with the GR value $\chi^2_\mathrm{GR}$.

\begin{figure}
   \centering
   \includegraphics[width=9cm]{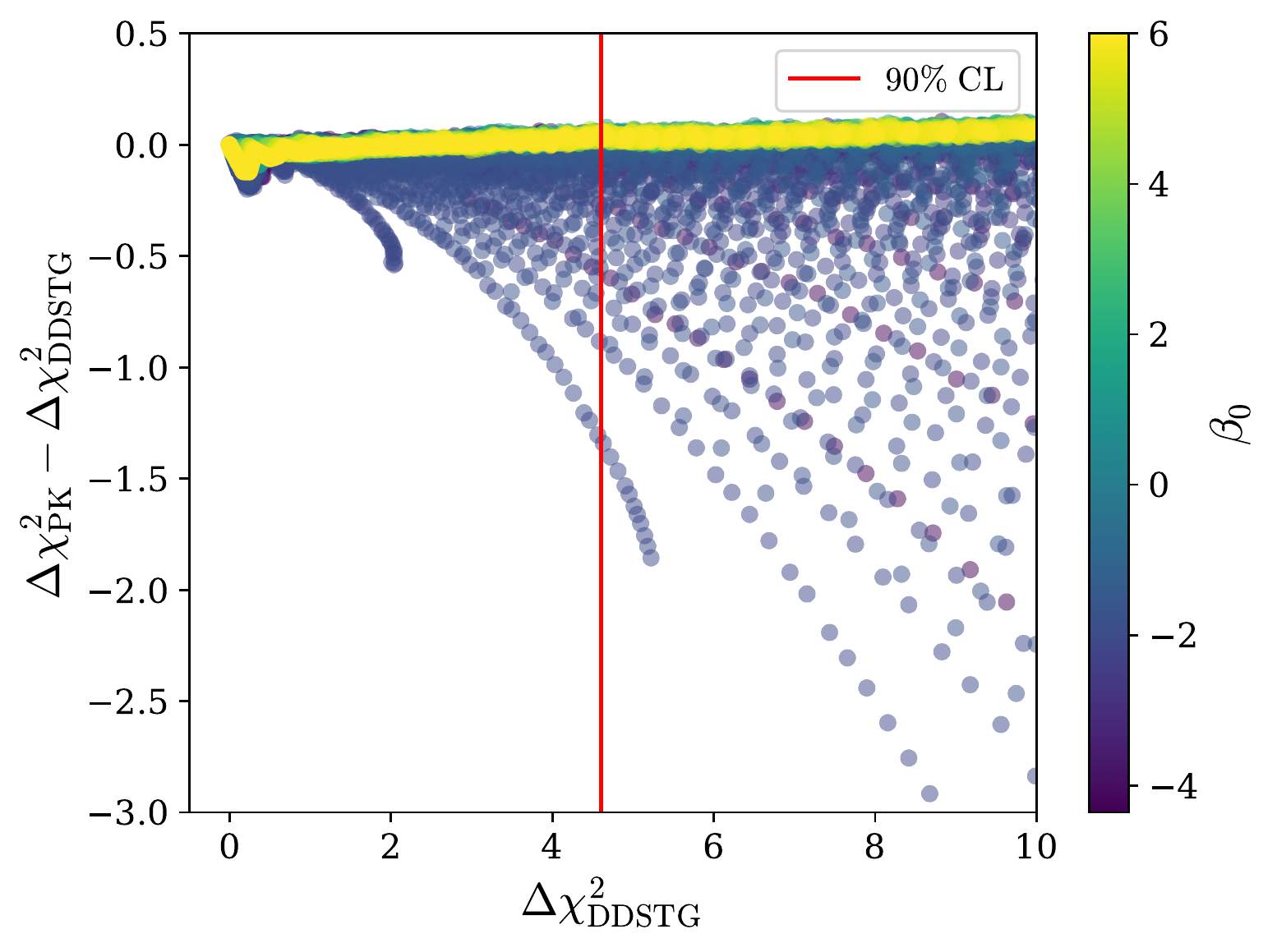}
      \caption{Comparison in the constraining power between the DDSTG model and the method based on measured PK parameters with the DD model. Both methods calculate $\chi^2$ values for a grid in $\{\alpha_0, \beta_0\}$ space, each point corresponds to a unique theory. $\beta_0$ values are presented by the colour, while $\alpha_0$ values cover the $[-10^{-1},-10^{-4}]$ range. Red vertical line corresponds to the $90\%$ CL limit, all points to the left are allowed by the test.}
         \label{Fig:J2222_DDSSTG_PK_comp}
   \end{figure}

Figure~\ref{Fig:J2222_DDSSTG_PK_comp} shows the comparison between the two methods. Each point on the plot presents a particular pair of $\{\alpha_0, \beta_0\}$ parameters. The analysis shows that, for PSR~J2222$-$0137, the DDSTG model produces higher or equal values of $\Delta\chi^2$ compared to the traditional ``PK method''. Despite the absence of any strong correlations between the PK parameters in PSR~J2222$-$0137, the new approach produces slightly more restrictive results for the area with negative $\beta_0$. For certain areas in $\{\alpha_0, \beta_0\}$ space the difference in the shifted $\Delta\chi^2 = \chi^2 - \chi^2_\mathrm{min}$ values becomes statistically significant when we calculate contours of $90\%$ confidence level limit ($\Delta \chi^2 \simeq 4.6$). 

\section{Predictions for PSR~J2222\texorpdfstring{$-$}{--}0137}
\label{sec:J2222_future}

Our next step is to estimate what enhancement in limits we can expect with future observations of PSR~J2222$-$0137. For this purpose we simulate fake TOAs for a set of radio observatories, assuming realistic timing precision estimated from real timing data.

\subsection{Simulated data-sets for FAST, MeerKAT and 3ERT}
{\renewcommand{\arraystretch}{1.5}
\begin{table}[t]
\caption{\label{tab:sim}Parameters of telescopes used in simulations.}
\begin{center}
\begin{tabular}{lccc}
\hline \hline
Telescope & $\diameter_{\mathrm{eff}}$ (m) & Bandwidth (MHz) & $\sigma_{\mathrm{TOA}}$ ($\mu$s) \\ \hline
Effelsberg & 100 & 200/400 & 2.67 \\
Nançay     & 94  & 512 & 2.02 \\
Lovell     & 76  & 400 & 3.66 \\
MeerKAT    & 108 & 856 & 0.84 \\ 
FAST       & 300 & 500 & 0.12 \\ \hline
\end{tabular}
\tablefoot{Effective diameter of telescopes, observing bandwidth, and TOA uncertainties of PSR~J2222$-$0137 used in the simulations. All information is based on the L-band data from real observations, and is scaled to 15-min integration over the full bandwidth.}
\end{center}
\end{table}}

The PSR~J2222$-$0137 timing data used up to this point are described in Section~\ref{sec:J2222_obs}. We will now describe our simulations, which show what we might be able to achieve in the near and foreseeable future with timing from this system.

We simulate TOAs spanning 10 years from 2021 to 2030 based on the current precision of the TOAs obtained with the 3ERT telescopes, as well as the TOA precisions from ongoing observations from MeerKAT and FAST. These simulations are conservative since they assume that there will be no improvement in the existing capabilities at these telescopes.

For TOAs from FAST, the radiometer noise reduces significantly thanks to its large collecting area, while the jitter noise becomes the primary limitation of timing precision. We find, however, that increasing the integration time to 15~min can largely reduce the jitter noise and eliminate its contribution to the timing precision (as it scales with the number of averaged pulses ($N_p$) as $\sigma_\mathrm{J}\propto 1/\sqrt{N_p}$ \citep{lk04}. Therefore, the median TOA uncertainty from 15-min TOAs are adopted in the simulation. 

Table~\ref{tab:sim} lists the telescopes assumed in our simulation, with the information on their effective diameter ($\diameter_{\mathrm{eff}}$), observing bandwidth, and TOA uncertainties at L-band. All TOA uncertainties are scaled to an integration time of 15~min over the full bandwidth. For each telescope, we assume one full orbit observation ($\sim$60 hours) per year, and split the observations into a monthly cadence, i.e. 5~hours per month\footnote{This may be unrealistic as telescopes may be oversubscribed. However, we found, e.g. for FAST, that with a more realistic assumption of 30 minutes per month, the uncertainty in the observed $\dot{P}_\mathrm{b}$ is only 5\% worse than observing 5 hours per month after 10-yr timing.}, to allow a good estimation of timing parallax (which requires a good coverage of Earth's orbit). This is important for the estimation of uncertainties in the external $\dot{P}_\mathrm{b}$ contributions shown in the next section.

The simulations are performed using the program developed in \cite{hkw2020}. First, we simulate TOAs based on the above assumptions, and add the TOAs from Effelsberg, Nan\c{c}ay, and Lovell telescopes together to be compared with MeerKAT and FAST. For 3ERT, the simulated TOAs are combined with the existing TOAs in \cite{gfg21}.
We then adjust the TOAs to perfectly match the timing parameters measured in \cite{gfg21}, and add a Gaussian white noise to each TOA based on its $\sigma_{\mathrm{TOA}}$. Finally, we fit for timing parameters and obtain their uncertainties, among which $\dot{P}_\mathrm{b}$ and timing parallax are of most important here. The whole process is done with the \textsc{TEMPO} DDSTG model.

\subsection{Contributions to the  uncertainty of \texorpdfstring{$\dot{P}_\mathrm{b}$}{PBDOT}}
\begin{figure}[t]
    \centering      \includegraphics[width=9cm]{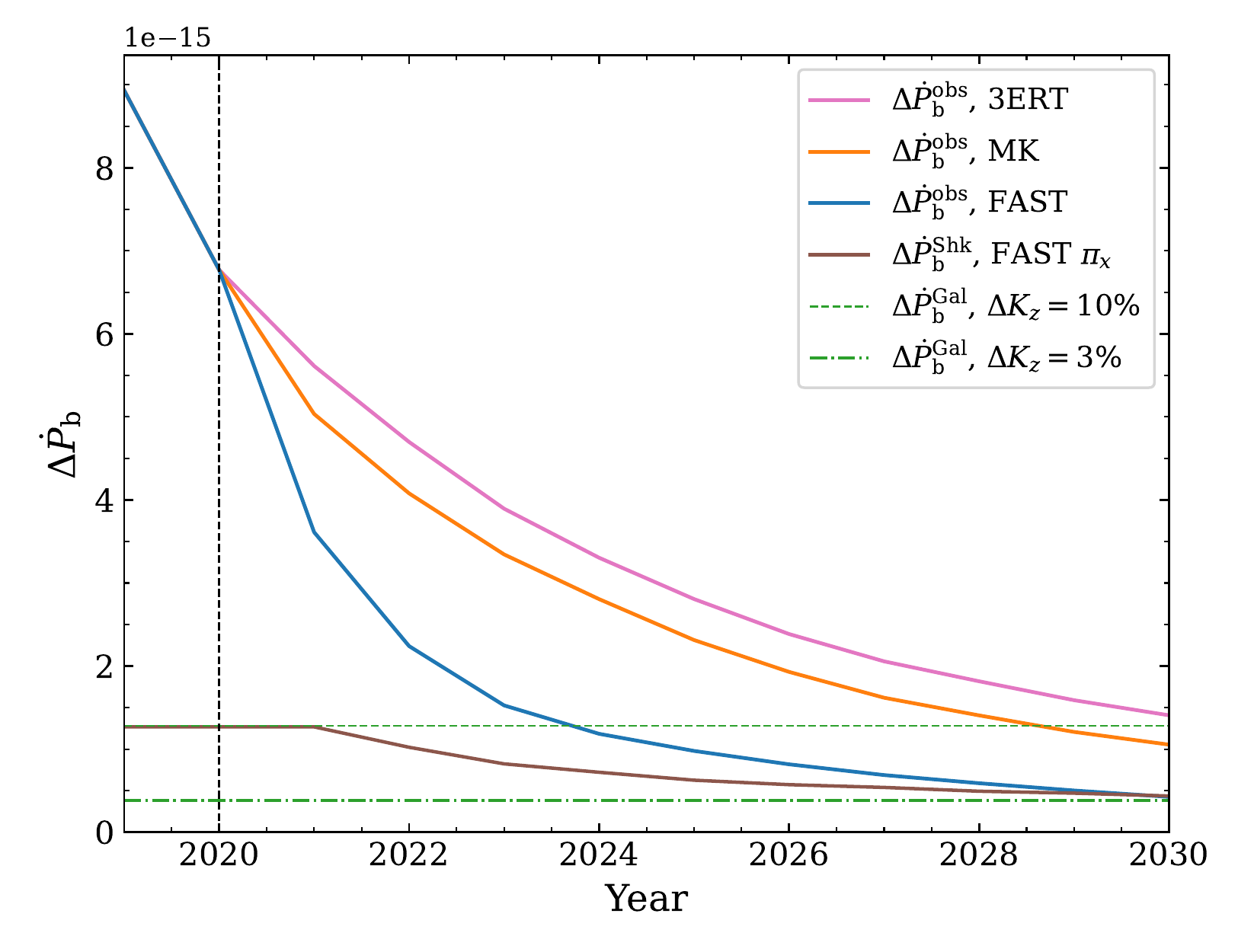}
    \caption{Comparison of different contributions in the uncertainty of $\dot{P}_\mathrm{b}$ for simulated data from 2021 to 2030. The pink, orange and blue lines show the uncertainty of $\dot{P}_\mathrm{b}$ using the simulated data from 3ERT (EFF, NC, LT), MeerKAT and FAST, respectively. The brown line indicates the uncertainty of $\dot{P}_\mathrm{b}^{\mathrm{Shk}}$ when using the timing parallax ($\pi_x$) measured from simulated FAST data. The green lines indicate the uncertainties of $\dot{P}_\mathrm{b}^{\mathrm{Gal}}$ when assuming 10\% (dashed) and 3\% (dash-dotted) uncertainties in the vertical component of Galactic acceleration $K_z$.}
    \label{Fig:d_PBDOT}
\end{figure}

The predicted uncertainties of $\dot{P}_\mathrm{b}$ are shown in Figure~\ref{Fig:d_PBDOT}, where the pink, orange, and blue lines show the improvement in time with the simulated data from 3ERT, MeerKAT, and FAST, respectively. As discussed in Section~\ref{subsec:PBDOT_uncertainties}, we also need to account for uncertainties from external effects, $\dot{P}_\mathrm{b}^{\mathrm{ext}}$. They are also expected to become more precise in the future because of anticipated improvements in the model for the Galactic gravitational potential, distance, and proper motion from future observations. Correspondingly improved values for $\dot{P}_\mathrm{b}^{\mathrm{Gal}}$ and $\dot{P}_\mathrm{b}^{\mathrm{Shk}}$ may then be estimated using Eqs.~\eqref{eq:pbdot_gal} and \eqref{eq:pbdot_shk}. For the Shklovskii effect, we find that it is mostly limited by the uncertainty in the distance, which comes from VLBI parallax or timing parallax measurement, whichever is better. With the simulated FAST data, the  measurements of timing parallax and proper motion improve quickly with time. In particular, the uncertainty of timing parallax will soon surpass the VLBI parallax and hence improve the distance measurement. The corresponding uncertainty in $\dot{P}_\mathrm{b}^{\mathrm{Shk}}$ will decrease with time as shown by the brown line in Figure~\ref{Fig:d_PBDOT}. This line is below the predicted uncertainties from observed $\dot{P}_\mathrm{b}$ most of the time, except at the very end of the simulation when compared to $\Delta\dot{P}_\mathrm{b}^\mathrm{obs}$ with FAST (blue line). In addition, future VLBI parallax measurements will likely be improved so that Shklovskii effect will not be a limiting factor for $\dot{P}_\mathrm{b}$. 

As for the contribution from the Galactic acceleration, a typical uncertainty in its vertical component ($\Delta K_z$) is about $10\%$, which contributes the most in $\Delta\dot{P}_\mathrm{b}^{\mathrm{Gal}}$, shown as the green dashed line in Figure~\ref{Fig:d_PBDOT}. 
With FAST, $\dot{P}_\mathrm{b}$ will then be limited by $\Delta K_z$ from 2024 onwards, if there is no improvement for this quantity. In fact, we find that $\Delta K_z$ needs to be improved to $\lesssim 3\%$ (see the green dash-dotted line) to not limit the precision of $\dot{P}_\mathrm{b}$ before 2030. The uncertainties in the Galactic potential do not limit the precision in this case. 
For the scope of further analysis, we assume that with future observations on pulsar timing, VLBI parallax, and Galactic acceleration, the precision of $\dot{P}_\mathrm{b}$ will not be limited by Shklovskii and Galactic effects. For Shklovskii contribution it is a reasonable assumption at least up to 2030. We also expect our knowledge about the Galactic potential to improve in the future especially in the proximity of the Solar System, thus we assume the improvement of $\Delta K_z$ for the selected very close pulsar to be $3\%$.

\subsection{Potential future constraints on DEF gravity from PSR~J2222--0137}

   \begin{figure}
   \centering
   \includegraphics[width=9cm]{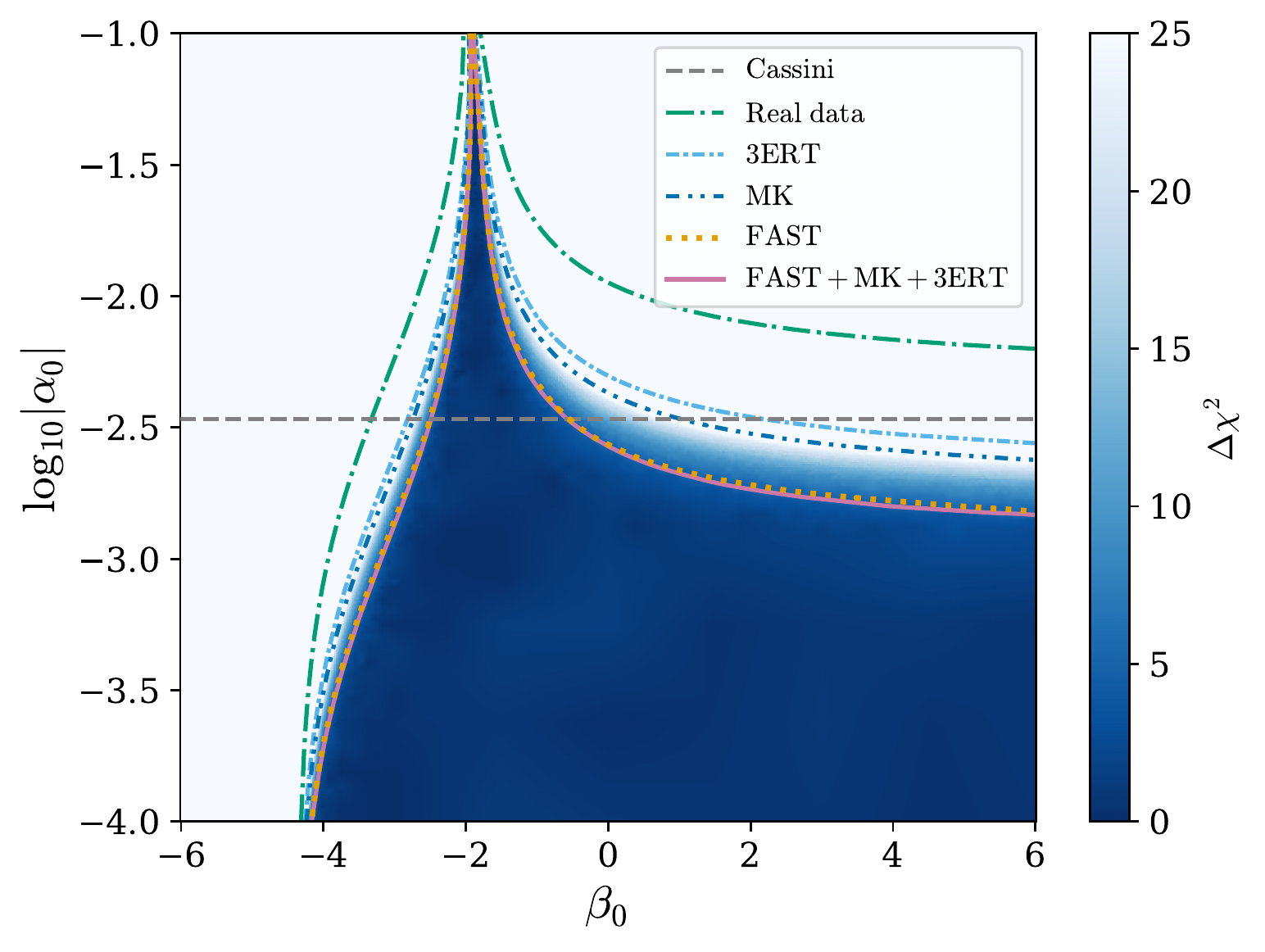}
      \caption{$\Delta \chi^2$ map for PSR~J2222$-$0137 from simulated data (2021--2030) combining different observatories and using MPA1 EOS. Assumed 1 orbit per year for FAST, MeerKAT, and 3ERT (EFF, NC, LT) based on their observation precision, appending to the existing data-set \citep{gfg21}. Each contour corresponds to $90\%$ CL limits on DEF gravity parameters from different data-sets, grey line depicts the limit from the Cassini mission.
              }
         \label{Fig:J2222_sim_chisqr}
   \end{figure}

We apply the same techniques as described in Section~\ref{sec:J2222_now} including the adaptive mesh refinement for contours. The map of $\Delta \chi^2$ for a combined simulated data is presented in Figure~\ref{Fig:J2222_sim_chisqr}. By 2030, we can expect a significant improvement in the limits for large positive $\beta_0$. This region is susceptible to dipolar gravitational emission. The significant improvement in $\dot{P}_\mathrm{b}$ measurement pushes the limit below the Cassini limit.

In Figure~\ref{Fig:J2222_sim_chisqr} we also present the comparison of constraining power for different observatories. The tightest constraints may be obtained for the combination of all three observatories. However, the precision of FAST is high enough to be significantly constraining on its own.

\section{Simulations for PSR-BH systems}
\label{sec:PSRBH}

   \begin{figure*}
   \centering
   \includegraphics[width=9cm]{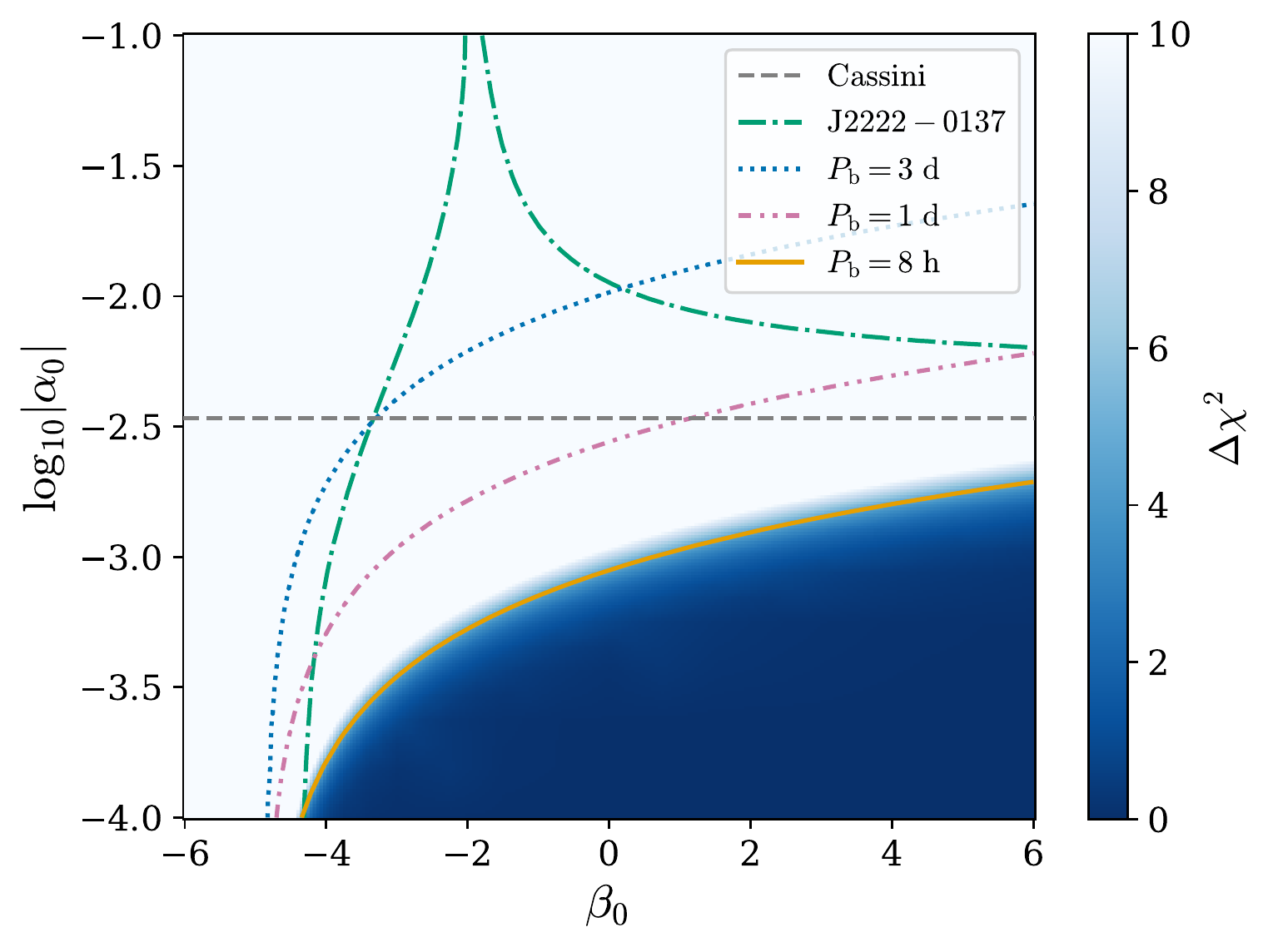}
    \includegraphics[width=9cm]{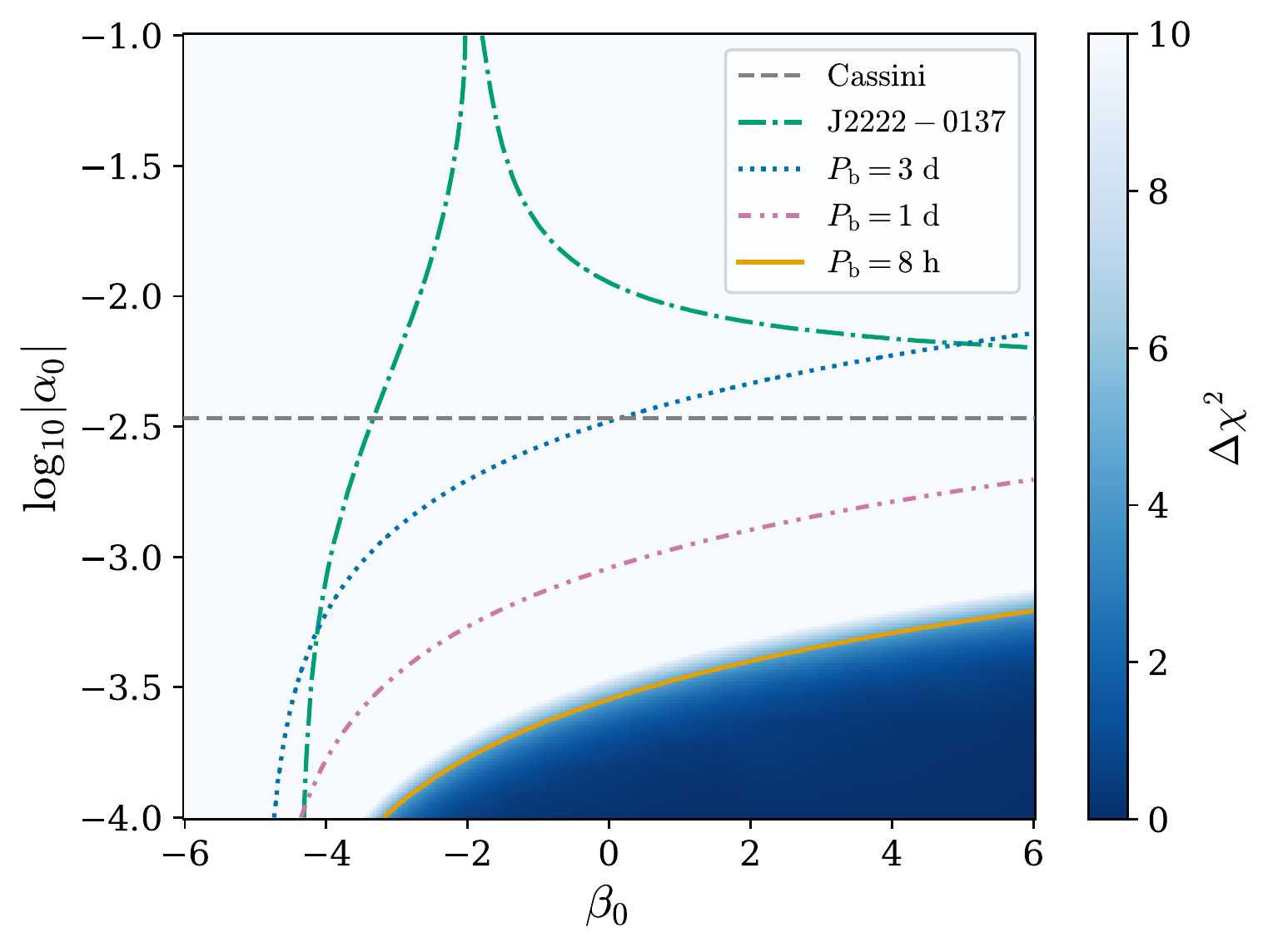}
      \caption{$\Delta \chi^2$ map in the DEF gravity parameter space from the simulated 5-yr timing data for three different PSR-BH systems. The left panel corresponds to $10\, \mu$s TOA uncertainty, the right panel to $1\, \mu$s. Solid lines show the $90\%$ CL limits for PSR-BH systems with different orbital periods: 3 d, 1 d and 8 h. The grey dashed line depicts the limit from the Cassini mission and green dash-dotted line is the current limit from PSR~J2222$-$0137 discussed in Sec.~\ref{sec:J2222_now} (see Fig.~\ref{Fig:J2222_chisqr}).}
      \label{Fig:PSRBH_chisqr_comp}
   \end{figure*}

As a further application of the new timing model, we investigate what limits on DEF gravity we can obtain from a  pulsar-black hole (PSR-BH) system. As mentioned above, our DDSTG implementation already allows to specify a BH as a companion, even though such a system has so far not been found. We apply the same technique used in Sections \ref{sec:J2222_now} and \ref{sec:J2222_future} to synthetic TOAs for three different hypothetical PSR-BH systems.

\subsection{Simulated FAST data-sets}

To investigate possible limits on DEF gravity from PSR-BH systems, we select presumed realistic parameters for simulating fake TOAs. These systems are assumed to comprise a $1.4\ M_{\odot}$ pulsar and a $10.0\ M_{\odot}$ BH in a highly eccentric orbit ($e=0.6$). As PSR-BH systems are more likely to reside in GCs (see the discussion in Section~\ref{sec:origin} and references therein), we assume that our hypothetical systems are located in the GC M15 and take the position (right ascension ``RA'' and declination ``DEC'') and dispersion measure (DM) of PSR~B2127+11C (M15C) for our simulations. To investigate how the limits on DEF gravity depend on the orbital period, we consider three cases with orbital periods of 3 d, 1 d, and 8 h, respectively. The selected parameters for PSR-BH systems are presented in Table \ref{tab:PSRBH_parameters}.

{\renewcommand{\arraystretch}{1.5}
\begin{table}
    \caption[]{Properties of pulsar-black hole systems taken for investigation.}
    \label{tab:PSRBH_parameters}
    \centering
    \begin{tabular}{ll}
        \hline\hline
        Parameters  & Values \\
        \hline
        RA ($\alpha$, J2000) & 21:20:01.2 \\
        DEC ($\delta$, J2000) & 12:10:38.2 \\
        DM\,(pc\,cm$^{-3})$ & $67.1$ \\
        Pulsar mass ($m_{\mathrm{p}}, M_{\odot}$) &  $1.4$ \\
        BH mass ($m_{\mathrm{BH}}, M_{\odot}$)    & $10.0$ \\
        Eccentricity ($e$)     & $0.6$\\
        Orbital period ($P_\mathrm{b}$, d) & \{$1/3$,  $1$, $3$\} \\
        TOA uncertainty ($\mu$s) & \{$1$, $10$\}\\
        Number of TOAs & $3144$ \\
        \hline
    \end{tabular}
\end{table}
}

We simulate data-sets with two different TOA uncertainties (for 15-min integration time): moderate $10\,\mu$s (which is typical for pulsars in GCs), and precise $1\,\mu$s. Taking these two, by an order of magnitude different TOA uncertainties helps to explore the dependence of the DEF gravity test on the TOA precision. We effectively assume a recycled millisecond pulsar, because they in general produce a better timing precision. We assume 6 hours of timing observation on these systems every two weeks, each session starts at a random orbital phase. All simulations cover a time span of 5~years and have the same number of TOAs ($n_\mathrm{TOA} = 3144$).

\subsection{How does a BH companion change the test of DEF gravity?}

Binary systems consisting of a pulsar and a white dwarf (PSR-WD) are particularly interesting for constraining STG theories due to their high asymmetry in compactness $(\alpha_{\mathrm{WD}} \simeq \alpha_0)$. The theory predicts a higher rate of orbital energy loss due to dipolar radiation from such asymmetric systems \citep{wil93,de96new}. However, generally, PSR-BH systems are expected to be even more asymmetric up to a certain positive $\beta_0$, as was pointed out by \cite{de98}. As a result of the no-scalar-hair theorem to BHs in STG one has \citep{haw72,de92}
\begin{equation}
    \alpha_{\mathrm{BH}} = 0,\quad \beta_{\mathrm{BH}} = 0 \, .
\end{equation}
This approximation is only valid for stationary BHs with an asymptotically flat spacetime and an asymptotically constant scalar field \citep{bcg+13}, which is not true anymore in the presence of a compact scalarised companion. However, a justification that it is still an excellent approximation can be found in \cite{lewk14}. The pulsar orbiting in an eccentric orbit will eventually induce a time-dependent scalar field at the location of the BH. This scalar field results in an induced effective scalar charge ($\alpha_{\mathrm{BH}}^{\mathrm{induced}}$) which, however, is totally negligible ($\lesssim 8 \times 10^{-14}\,\alpha_{\rm p}$; cf.\ Eq.~(50) in \citealt{lewk14}).

The absence of scalar charges for the BH results in simplified relations of the PK parameters. The main consequence of a BH presence is that all PK parameters except $\dot{P}_\mathrm{b}$ are identical to their GR expressions with an appropriate rescaling of the masses. In the GW damping sector, all the multipoles, e.g., monopolar, dipolar and quadrupolar modes, are still present. Thus PSR-BH systems are susceptible but only to the test of GW damping via $\dot{P}_\mathrm{b}$ \citep{de92,mw13,lewk14}.

\subsection{Results}

{\renewcommand{\arraystretch}{1.5}
\begin{table}
    \caption[]{\label{tab:PSRBH_Pbdot_uncertainties}Orbital period properties of simulated PSR-BH systems.}
    \centering
    \begin{tabular}{c|c|cc}
        \hline\hline
        $P_\mathrm{b}$ (d) & $\dot{P}_\mathrm{b}^\mathrm{GR}$ ($10^{-12}$ s s$^{-1}$) & $\Delta\dot{P}_\mathrm{b}, 10$\,$\mu$s& $\Delta\dot{P}_\mathrm{b}, 1$\,$\mu$s\\
        \hline
        $3$  & $-0.22220$ & $0.08031$ & $0.00827$ \\
        $1$   & $-1.38661$ & $0.01723$ & $0.00185$ \\
        $1/3$ & $-8.65279$ & $0.00532$ & $0.00055$ \\
        \hline
    \end{tabular}
    \tablefoot{The rate of the orbital period change in GR ($\dot{P}_\mathrm{b}^\mathrm{GR}$) for selected PSR-BH systems and corresponding uncertainty in $\dot{P}_\mathrm{b}$ measurement for three different orbital periods and two TOA uncertainties. The values of $\dot{P}_{\mathrm{b,GR}}$ and $\Delta\dot{P}_\mathrm{b}$ are given in $10^{-12}$ s s$^{-1}$.}
\end{table}
}

The limits obtained from the DDSTG model for the simulated PSR-BH systems are shown in Figure~\ref{Fig:PSRBH_chisqr_comp}. For the tests presented here we do not include any potential uncertainty in the observed $\dot{P}_\mathrm{b}^\mathrm{obs}$ value due to the external effects. The restrictive power of the test increases for more relativistic systems with shorter orbital periods. To place new limits on DEF gravity with a moderate TOA precision of $10\,\mu$s the orbital period should be a fraction of a day. 

Another way to improve the restrictive power is to increase the precision of the TOAs (compare left and right panels of Figure~\ref{Fig:PSRBH_chisqr_comp}). The obtained limits strongly depend on the accuracy of $\dot{P}_\mathrm{b}$ measurement as most of the deviations from GR come from the predicted dipolar GW emission. Table~\ref{tab:PSRBH_Pbdot_uncertainties} shows the predicted GR values of $\dot{P}_\mathrm{b}$ and compares them to the corresponding $\Delta\dot{P}_\mathrm{b}$ uncertainties. The lower uncertainty and the higher the predicted value the better the test. 

We expect a dramatic improvement of limits on DEF gravity from relativistic PSR-BH system. Due to higher asymmetry, the test is extremely sensitive to the precision of $\dot{P}_\mathrm{b}$ measurement. To answer the possibility of obtaining accurate enough TOAs, we have to argue where we can find such a system.

\subsection{Pulsar-black hole systems origin}
\label{sec:origin}
A binary pulsar system with a stellar-mass BH may originate from several different evolutionary scenarios. The first way is a standard evolution of a massive binary system \citep{vt03}. It results in a PSR-BH system with a young slow spin-period ($\sim0.1-1$ s) pulsar and a wide, eccentric orbit. The second path is a reversal mechanism \citep{spn04} taking place under a specific set of circumstances. The pulsar is formed first and later spun-up by accretion during the red-giant phase of the companion \citep{ppp05}, later becomes a BH. This is similar to the origin of PSR~J1141$-$6545, a system where a massive WD star formed before a more massive NS \citep{ts00a}. 
The reversal mechanism may give a more desirable result of a recycled pulsar in a system with a BH, because recycled pulsars have generally more precise timing and a more stable rotation than the slow ``normal'' pulsars \citep{vbc+09}.

Moreover, the third possible way to form a PSR-BH system is a multiple body encounter, happening in regions of high stellar density, e.g., GCs and the Galactic Centre region \citep{vbc+09,csc14}. 
Such encounters are the reason why there are $\sim 10^3$ times more low-mass X-ray binaries (LMXBs) per unit stellar mass in GCs than in the Galactic disk \citep{cla75}, which results in a similarly enhanced proportion of millisecond pulsars (MSPs). 
In these encounters, an old, inactive NS approaches a main sequence (MS) binary so closely that a chaotic interaction ensues. The most likely result of such an interaction is that the two most massive objects (in this case the old, recycled NS and the more massive MS star) will form a more compact binary system, with the lighter MS star component being ejected at high velocity (see review by \citealt{phi92}). That MS will evolve, fill its Roche lobe, and start transferring matter onto the NS, which is spun up in the process, this is the aforementioned LMXB stage.

If left undisturbed, many of these then evolve into binary MSP systems. A consequence of this is that GCs not only have many MSPs, but that the number of MSPs in each cluster appears to be roughly proportional to its rate of stellar encounters ($\Gamma$) \citep{vh87}.

However, in some GCs - especially those with collapsed cores - stellar densities are so high that only few binaries evolve without being disturbed; these GCs have a large interaction rate {\it per binary} ($\gamma_{\rm GC}$) \citep{vf14}. This means that, even after being recycled, an MSP has a high probability of undergoing further (``secondary'')  exchanges.

In such exchanges, an incoming massive star interacts chaotically with the components of either a LMXB or a MSP binary. Again, the least massive object is the most likely to be ejected, in this case that will be the low-mass star that recycled (or was still recycling) the NS. A new binary will form, consisting of the NS and the intruding massive star. If the latter is degenerate, then the system will not undergo accretion and the orbital circularisation that comes with it, but will keep the orbital eccentricity it acquired after formation.

Several such obvious products of secondary exchange encounters have been found in GCs, invariably with a large $\gamma_{\rm GC}$: PSR~B2127+11C, in the core-collapsed M15 GC \citep{pakw91,jcj+06}, PSR~J0514$-$4002A, in NGC~1851 \citep{fgri04,rfgr19} (recently two additional such systems, D and E,  were found in the same GC, see \citealt{rfg+22}), PSR~J1807$-$2500B, in the core-collapsed NGC~6544 \citep{lfrj12}, PSR~J835$-$3259A, in the core-collapsed NGC 6652 \citep{drk+15} and PSR~J1823$-$3021G, in the core-collapsed NGC~6624 \citep{rgf+21}. These discoveries suggest (but by no means assure us) of the possibility that similar secondary exchange encounters might produce MSP binaries with even more massive companions, such as BHs. As for the massive MSP binaries above, such MSP-BH systems would be preferentially produced in the GCs with large $\gamma_{\rm GC}$; this is one of the reasons why they are targeted by the MeerKAT TRAPUM survey \citep{rgf+21}.

These secondary exchange products are that their orbital periods vary between 8 h in the case of B2127+11C and 18.8 d in the case of PSR~J0514$-$4002A, their orbital eccentricities are between 0.38 and 0.90. Thus the simulated PSR-PH systems listed in Table~\ref{tab:PSRBH_parameters} have realistic orbital periods and eccentricities.

\subsection{Influence of a globular cluster origin}

For the pulsars observed in GCs, the derivatives of the spin period, $P$, and in the case of binary pulsars, the derivative of the orbital period, $\dot{P}_\mathrm{b}$ are contaminated (and in most cases dominated) by the line-of-sight component of the acceleration of the pulsar (or binary) in the gravitational potential of the GC ($a_{\mathrm{GC}}$); this enters Eq.~\eqref{eq:pb_obs} as an additional term that is similar to $\dot{P}_\mathrm{b}^{\mathrm{Gal}}$:
\begin{equation}
\dot{P}_\mathrm{b}^\mathrm{GC} = \frac{a_{\mathrm{GC}}}{c} P_{\mathrm{b}} \, .
\end{equation}
For pulsars in GCs, the only radiative test done to date was with PSR~B2127+11C \citep{jcj+06}. The orbital decay observed in this system is $-3.95 \pm 0.13 \times 10^{-12}\,{\rm s\,s^{-1}}$, which is within $\sim 3 \%$ of the predicted value. That test, however, is limited by the fact that the maximum value for $|a_{\mathrm{GC}}/c| \sim 6 \times 10^{-18} \, \rm s^{-1}$ \citep{phi93}, i.e., in this case the maximum value for $|\dot{P}_\mathrm{b}^\mathrm{GC}| \sim 0.17 \times 10^{-12}\,{ \rm s\,s^{-1}}$. This is of the same order as their measurement precision, which means that this radiative test cannot be improved, unless one could measure the acceleration of the pulsar independently. The situation would be much worse if PSR~B2127+11C were closer to the cluster centre, where accelerations are much larger. For instance, PSR~B2127+11A, a couple of arcseconds from the centre, has $P \, = \, 0.1106\,$s and $\dot{P} \, = \, -2.107 \times 10^{-17}$; this implies that $|a_{\mathrm{GC}}/c | > 1.9 \times 10^{-16} \, \rm s^{-1}$ and thus $|\dot{P}_\mathrm{b}^\mathrm{GC}| > 5.5 \times 10^{-12} \, \rm s\,s^{-1}$, a value larger than the orbital decay predicted by GR.

For the three PSR-BH systems in Table~\ref{tab:PSRBH_Pbdot_uncertainties}, the situation would be similar. If they were at the locations of PSR~B2127+11C, their values of $P_{\mathrm{b}}$ would imply that the maximum values for $\dot{P}_\mathrm{b}^\mathrm{GC}$ would be 9, 3 and 1 times larger respectively, i.e., $1.53$, $0.51$ and $0.17 \times 10^{-12}\, \rm s\, s^{-1}$. This corresponds to, respectively, $\sim 6.9$, $\sim 0.37$ and 0.02 times the values of $\dot{P}_\mathrm{b}^\mathrm{GR}$ in Table~\ref{tab:PSRBH_Pbdot_uncertainties}. That means that, for the latter system, a $\sim 50$-$\sigma$ test of the radiative properties of a PSR-BH system would be possible. These numbers illustrate the immense advantage of an increasingly shorter $P_{\mathrm{b}}$, either for GW tests in GCs or in the Galaxy: on one hand, the $\dot{P}_\mathrm{b}^\mathrm{GR}$ increases with $P_{\mathrm{b}}^{-5/3}$, the ``polluting'' part $\dot{P}_\mathrm{b}^\mathrm{GC}$ decreases as $P_{\mathrm{b}}$. Thus, the significance of a particular GW test grows, everything else being identical, with $P_{\mathrm{b}}^{-8/3}$.

However, if the 8-h PSR-BH system is placed at the location of PSR~B2127+11A, the test would lose its significance almost entirely. Fortunately, even in this case, we can get a firm upper limit for $|\dot{P}_\mathrm{b}^\mathrm{GR}|$; the reason is that we also measure the spin period derivative. That is also affected by the acceleration in the cluster and by other terms, such as the Shklovskii effect. Adding the equations for $\dot{P}$ and $\dot{P}_{\mathrm{b}}$, we obtain:
\begin{equation}
\dot{P}_\mathrm{b}^\mathrm{GR} = \dot{P}_\mathrm{b}^\mathrm{obs} - P_{\mathrm{b}} \left( \frac{\dot{P}}{P} \right)^\mathrm{obs} + \frac{1}{2 \tau_c} P_{\mathrm{b}} \, .
\end{equation}
Note that all the terms on the right can be measured precisely, except for the characteristic age of the pulsar, $\tau_c$, which cannot be measured independently. Therefore, the last term then quantifies the uncertainty of the test, which is, again, proportional to $P_{\mathrm{b}}$, which implies that the significance of the test is again proportional to $P_{\mathrm{b}}^{-8/3}$.

However, that last term is necessarily positive. If we assume the pulsar is extremely old, then this term will be very small and we obtain, from the other two terms, a hard lower limit for $\dot{P}_\mathrm{b}^\mathrm{GR}$. Since the latter is negative, this represents a hard upper limit of its magnitude. This means that such a test could still, in principle, falsify GR.
An upper limit on the last term can be obtained from the lowest likely value for $\tau_c$; for most MSPs $\tau_c \, < \, 1\,$Gyr. This means that, for the 8-h PSR-BH system, this unknown term would be at most $0.46 \times 10^{-12}\, \rm s\, s^{-1}$, implying a $\sim 20$-$\sigma$ test of GR. Thus, this test will be the more significant the larger $\tau_c$ is compared to the orbital decay timescale for the binary.

Despite such possible mitigation, it is clear that the location in a GC always degrades the quality of radiative tests. For the 8-h PSR-BH system discussed above, significances of 50 and 20 $\sigma$ in the measurement of $\dot{P}_{\rm b}$ represent a significant degradation relative to the tests listed in Table~\ref{tab:PSRBH_Pbdot_uncertainties}, where for a 8-h PSR-BH system timed with 10$\mu$s the significance of the $\dot{P}_{\rm b}$ measurement is larger than 1600.

\section{DDSTG and other gravity theories}
\label{sec:other_theories}

The approach developed in this work is not restricted to DEF gravity. It can straightforwardly be extended to investigate a larger set of alternative gravity theories. Broadly speaking, every theory that maps to the DD phenomenological model \citep{dd86,dt92} can be put into the DDSTG framework with a new implementation of appropriate PK formulae. The DD model is a quasi-Keplerian solution in the first post-Newtonian approximation to the dynamics of a 2-body system within the modified Einstein-Infeld-Hoffmann (mEIH) framework \citep{dt92,wil93,wil18}. The mEIH formalism covers a large set of fully-conservative gravity theories without the ``Whitehead term'' in the post-Newtonian limit. Later it was extended by \cite{wil18b} from fully conservative to semi-conservative theories of gravity. However, this extension requires additional terms in mEIH Lagrangian, which are not accounted for in the DD model.

The DDSTG model can be directly applied to a certain class of gravity theories without a modification of the PK parameter equations. For example, DEF gravity is a specific case of a massless mono-scalar tensor gravity theory with a particular expression of the conformal coupling $A(\varphi)$. DDSTG covers STG theories with any conformal coupling function depending on two arbitrary parameters $\{\alpha_0, \beta_0\}$ and a single massless scalar field. To work, the model only requires pre-calculated gravitational form-factors for the new coupling function. In case of a more complex coupling function depending on a higher number of parameters the \textsc{TEMPO} code needs to be adapted.

Recently, Mendes and Ortiz \citep[MO,][]{mo16} introduced an extension of DEF gravity. MO gravity is an example of a theory that can be easily incorporated into the DDSTG model. Its difference from DEF gravity is in the form of the conformal coupling
\begin{equation}
    A(\varphi) = \left[\cosh\left(\sqrt{3}\beta_0\varphi\right)\right]^{1/(3\beta_0)} \, ,
\end{equation}
where $\beta_0$ is a free parameter. The second parameter $\alpha_0$ is hidden in MO gravity to the scalar field at infinity. MO theory received attention in recent years and was originally introduced as an analytical approximation to a more fundamental theory, where the action includes quadratic terms of the scalar field coupled to curvature. The developed framework can therefore be straightforwardly extended to MO gravity without the change of the \textsc{TEMPO} implementation.

\section{DDSTG and a time-varying gravitational constant}
\label{sec:gdot}

Another interesting extension of the DDSTG model---planned for a future release---is the inclusion of the effects of a temporal variation of the gravitational constant ($G$). A time-evolving asymptotic scalar field ($\varphi_0$) of a gravitating system, generally, leads to a temporal variation of the local gravitational constant. In DEF gravity, such a change of the Newtonian gravitational constant as measured in the Solar System reads 
\begin{equation}
    \frac{\dot{G}_\mathrm{Cav}}{G_\mathrm{Cav}} = 2\left(1 + \frac{\beta_0}{1 + \alpha_0^2}\right)\alpha_0\dot\varphi_0 \,
\end{equation}
\citep[see e.g.][]{uza11}. Generally, one expects a temporal change in $\varphi_0$ to arise from the expansion of the universe and $\dot\varphi_0$ to result from a cosmological model based on DEF gravity \citep[see e.g.][]{Damour:1993id}. However, as part of a more agnostic approach, $\dot\varphi_0$ can be treated as an additional, independent timing parameter.

The main impact on the orbital motion of a binary system that arises from a time-varying gravitational constant is a secular change in the orbital period \citep{dgt88}. For two weakly self-gravitating masses, one simply has $\dot{P}_\mathrm{b}^{\dot{G}}/P_\mathrm{b} = -2\dot{G}/G$. However, for binary pulsar systems, this simple expression needs to be extended by body-dependent contributions \citep{nor90,nor93b}. One then has
\begin{equation}
    \frac{\dot{P}_\mathrm{b}^{\dot{G}}}{P_\mathrm{b}} = - 2 \left( \frac{\dot{G}_\mathrm{Cav}}{G_\mathrm{Cav}}\right) \mathcal{F}_{AB} \, ,
\end{equation}
where the factor $\mathcal{F}_{AB}$ accounts for all the corrections related to the strong gravitational fields of the pulsar and its companion, if the latter is also a NS. More specifically, following \cite{nor93b}, $\mathcal{F}_{AB}$ accounts for a change in the body-dependent part in the effective gravitational constant $G_{AB}$ as well as a change in the masses resulting directly from $\dot\varphi_0$.\footnote{Detailed expressions for $\mathcal{F}_{AB}$ will be given in a future publication. See also \cite{wex14}.}

It has been demonstrated by \cite{wex14} that, depending on the parameter space, pulsar mass and EOS, strong-field effects can considerably enhance the effect of a time-varying gravitational constant, i.e. $\mathcal{F}_{AB} \gg 1$. Consequently, accounting for $\dot{P}_\mathrm{b}^{\dot{G}}$ in binary pulsar tests not only provides an independent test for a varying gravitational constant but also probes strong-field aspects related to a time-varying gravitational constant.

\section{Discussion and conclusions}
\label{sec:conclusions}

In this work, we developed a new, improved approach for testing STG. We examined a specific class of STG theories known as ``DEF gravity''. This approach is based on a new timing model, called the DDSTG model, which is an extension of the DD model to work within DEF gravity. Analysis of pulsar timing data with this model overcomes some of the problems of conventional methods, which we have discussed in detail. The DDSTG timing model uses theoretical predictions for PK parameters in DEF gravity and therefore uses a minimal set of binary parameters. For that reason, it accounts for all the possible correlations between these parameters. All the information from the observational data is used to provide the most reliable tests of an alternative theory, directly without intermediate steps with phenomenological parameters of the DD model.

As a demonstration of the DDSTG model, we applied it to the most recently published timing data of the binary pulsar PSR~J2222$-$0137 described and used by \citet{gfg21}. This system is of great importance for testing alternative gravity theories because it is very close to us (resulting in the most precise VLBI distance for any pulsar), has precise timing and shows a set of well measured relativistic effects. The system has a high asymmetry in the compactness between the components: it comprises a massive NS ($m_\mathrm{NS} \sim 1.82 M_\odot$) and a massive WD ($m_\mathrm{WD} \sim 1.31 M_\odot$). This high asymmetry results, for some areas in the DEF gravity parameter space, in the prediction of a very strong dipolar GW contribution to the rate of orbital decay ($\dot{P}_\mathrm{b}^D$). The non-detection of dipolar GWs in this system  is used to constrain the DEF gravity parameter space. Moreover the mass of the pulsar lies in the ``scalarisation gap'' ($m_\mathrm{NS} \gtrsim 1.5 M_\odot$); this means that strict limits on the occurrence of spontaneous scalarisation, a highly non-linear phenomenon, can be placed \citep{zfk+22}.

The results from the new method confirmed and improved the existing limits on DEF gravity parameters from this system. The DDSTG model appeared to be more constraining in the area near spontaneous scalarisation ($\beta_0 \lesssim -4.0$) when compared to the commonly used PK method.  It suggests that the combination of the DDSTG model with an EOS agnostic approach can improve limits placed on the spontaneous scalarisation.

Moreover, we applied the DDSTG timing model to the simulated TOA for PSR~J2222$-$0137 covering the period of 2021--2030 to see what improvement we can expect from that system in the future. The mock timing data-sets were simulated for several large observatories FAST, 3ERT (EFF+NC+LT), MeerKAT with TOAs uncertainties based on the real timing data. We discussed the future importance of kinematic contributions (Shklovskii and Galactic) to $\dot{P}_\mathrm{b}$, and consequently to the precision of these tests. The main limiting factor to $\dot{P}_\mathrm{b}$ comes from the uncertainty in the Galactic contribution ($\dot{P}_\mathrm{b}^\mathrm{Gal}$). In particular, the limit comes from the current uncertainty in the vertical component of the Galactic acceleration ($\Delta K_z$). Our analysis predicts that this limitation will disappear if the current uncertainty of $\Delta K_z \sim 10 \%$ can be improved to $\Delta K_z \lesssim 3\%$, which is likely to be the case in the future with improvement of models for the gravitational potential of our Galaxy. Future observations are expected to significantly improve the limits on DEF gravity, especially with the use of FAST data. 

One of the most promising systems for testing gravity, which we hope to have in the near future, are binary pulsar-black hole systems (PSR-BH). We simulated artificial timing data-sets for three eccentric PSR-BH systems with reasonable orbital parameters for three different orbital periods. The results of applying the DDSTG model to the simulated PSR-BH data strongly depended on the precision of the $\dot{P}_\mathrm{b}$ measurement. We briefly discussed possible evolution scenarios leading to the formation of the PSR-BH system, such as GC origin and reversal mechanism. Depending on the place of origin, there might be issues in obtaining a precise intrinsic $\dot{P}_\mathrm{b}$, i.e. accounting for the contamination of $\dot{P}_\mathrm{b}$ from a kinematic contribution due to the acceleration of the system in the gravitational field of the GC. Depending on the timing precision and orbital properties, PSR-BH can place stringent limits on DEF gravity.

In the future, the DDSTG model can be applied to a range of different binary pulsar systems to improve the limits on DEF gravity. We expect especially interesting results from PSR~J1141$-$6545 where DDSTG is expected to be superior to standard approaches \citep{vvk19}. PSR~J1141$-$6545 is an asymmetric PSR-WD system in 4.7 hours orbit with significant spin-orbit coupling due to the fast rotating WD. Due to the spin-orbit coupling the system shows a change of the projected semi-major axis $\dot{x}^\mathrm{SO}$ which has a strong 
correlation with the time dilation parameter $\gamma$. The latter parameter is caused by the precession which cannot be calculated, because we do not know the exact spin properties of the WD. Moreover, PSR~J1141$-$6545 shows a weak Shapiro delay in the timing data which the DDSTG model can fully exploit resulting in a significant improvement of the test. The high asymmetry in the compactness between the components ($\massp\sim 1.26 M_\odot$ NS and $\mc\sim 1.02 M_\odot$ WD) makes this system a perfect tool for radiative tests of gravity. We expect the DDSTG model to be of particular advantage because it accounts for both possible correlations and weak relativistic effects (Venkatraman Krishnan et al., in prep.).

Another perspective system to perform DEF gravity tests with the DDSTG model is the Double Pulsar PSR~0737$-$3039A which shows the largest number of PK parameters in the timing data \citep{ksm21}. The system consists of two radio pulsars with masses of $\massp \simeq 1.34\ M_\odot$ and $\mc \simeq 1.25\ M_\odot$. Properties of NSs in DEF gravity (gravitational form-factors) strongly depend on the choice of the EOS, which in turn affects the test. Thus to put reliable limits on DEF gravity from PSR~0737$-$3039A independent of a choice of EOS, one must perform an EOS-agnostic analysis. EOS-agnostic test means that it is performed with a set of EOSs which are diverse in their properties (see \citealt{vcf+20}, who used such an EOS-agnostic approach to constrain DEF gravity with a pulsar in a stellar triple system). With this paper we included a set of 11 EOSs varying from soft to stiff (see Appendix~\ref{App:pre-calculated_grids}) which can be used for such agnostic tests in the future.


\begin{acknowledgements}

We thank Lijing Shao for carefully reading the manuscript and giving important comments. We further thank Thomas Tauris and Selma de Mink for useful discussions and suggestions. AB and HH are members of the International Max Planck Research School for Astronomy and Astrophysics at the Universities of Bonn and Cologne. HH acknowledges the support by the Max-Planck Society as part of the ``LEGACY'' collaboration with the Chinese Academy of Sciences on low-frequency gravitational wave astronomy. This study is partly based on observations with the 100-m telescope of the MPIfR (Max-Planck-Institut f\"ur Radioastronomie) at Effelsberg. The Nan\c{c}ay Radio Observatory is operated by the Paris Observatory, associated with the French Centre National de la Recherche Scientifique (CNRS). LG, IC, and GT acknowledge financial support from the ``Programme National Gravitation, Références, Astronomie, Métrologie (PNGRAM)'' of CNRS/INSU, France. 
J. W. McKee gratefully acknowledges support by the Natural Sciences and Engineering Research Council of Canada (NSERC), [funding reference \#CITA 490888-16]. This publication made use of open source libraries including \textsc{DifferentialEquations.jl}\textsuperscript{\ref{url_diffeq}} \citep{DifferentialEquations.jl-2017}, \textsc{Optim.jl} \citep{mogensen2018optim}, \textsc{ForwardDiff.jl} \citep{RevelsLubinPapamarkou2016}, \textsc{Matplotlib} \citep{hun07}, along with a pulsar analysis package \textsc{TEMPO}\textsuperscript{\ref{url_tempo}} \citep{nds15}.
\end{acknowledgements}


\bibliographystyle{aa}
\bibliography{journals,modrefs,psrrefs,crossrefs,newrefs}


\appendix

\section{Details on calculating grids of NSs} 
\label{App:details_on_grids}

We select slowly rotating axisymmetric approximation of NSs following \citet{de96new}. The internal structure of such a NS in DEF gravity is described by a set of 8 ordinary differential equations depending on the radial variable ($r$). To calculate a single NS the program solves the system of ODEs by applying appropriate boundary conditions at the centre of a NS and spatial infinity. We use a shooting technique to match initial internal parameters with the external structure of the spacetime. The differential equations are solved by means of the Julia library \textsc{DifferentialEquations.jl}\footnote{\label{url_diffeq}\url{https://diffeq.sciml.ai}} \citep{DifferentialEquations.jl-2017}. 

A particular solution is determined by the choice of two theory parameters $\{\alpha_0, \beta_0\}$, the EOS and the central pressure $p_c$. The central value of the scalar field $\varphi_c$ is determined by the shooting procedure to correspond to the scalar field value at spatial infinity $\varphi_0 = \varphi(r=\infty)$, which in our case is, without loss of generality, set to zero. Thus $\varphi_c$ is not an arbitrary parameter. The central density ($p_c$) uniquely corresponds to the gravitational mass $m_A$ for most masses of interest and fixed theory parameters $\{\alpha_0, \beta_0\}$. In general, this is not the case --- there may be several stable solutions for a fixed mass in the area of strong scalarisation. However, multiple solutions happen for large NS masses and very negative $\beta_0 < -4.5$ which is already ruled out and thus of no interest.

In the recent papers devoted to the calculation of NSs in scalar-tensor gravity \citep{and19}, the gravitational form-factors from Eq.~\eqref{eq:scalar_charges} (they are also often called ``scalar charges'') are calculated using numerical differentiation. Numerical differentiation is performed on a grid of dependent variables. The precision of this approach is determined by two factors: a) the error in the calculation of the desired quantity and b) the error due to the numerical differentiation formula. The first error depends on the precision of the numerical integration of the structure equations. The second error strongly depends on the fineness of the grid on which the derivatives are calculated. If the grid is too sparse the formula such as the central difference formula $f'(x) \simeq \left[f(x+\delta x) - f(x-\delta x)\right]/\left[2\delta x\right]$ is not accurate enough. On the other hand, if the grid is too fine, the error arises because of the subtraction of two close numbers in the computer memory. The total relative errors for calculating gravitational form-factors usually lie in the range of $\delta_{\mathrm{rel}} \sim 10^0-10^{-5}$. In our work, we propose a more precise way to calculate all the quantities.

\subsection{Automatic Differentiation}
\label{App:AutoDiff}

In our program, we utilise the Automatic Differentiation (AutoDiff) technique. It is a special technique to calculate derivatives, when the algorithm knows the exact expressions for derivatives of all elementary functions and uses the chain rule to unfold complex derivatives. A review on the AutoDiff can be found, for instance, in \citet{mar18}. We use an implementation of the AutoDiff technique in Julia from ForwardDiff.jl library, which is a part of a broad JuliaDiff framework \citep{RevelsLubinPapamarkou2016}. AutoDiff allows calculating complex derivatives of quantities with respect to their arguments with the machine-precision  (for Double Float numbers the corresponding precision is $\delta \sim 2\times10^{-16}$ on 64-bit systems). It is done by internal implementation of sophisticated arithmetic on a special type of numbers (dual numbers) in the programming language. A derivative is calculated exactly in the desired point and does not require points in the vicinity. The precision of the method is constant and thus does not depend anymore on the fineness of the grid.

Automatic differentiation enables us to simultaneously calculate both the function $F = F(x,y)$ and its derivatives with respect to arguments $\frac{\partial F}{\partial x}, \frac{\partial F}{\partial y}$.  It takes twice as long as calculating a mere function itself and both are done with the machine-precision . To calculate a complex derivative, then the arguments are also functions $B = B(x,y),\ C = C(x,y)$ we apply the chain rule:
\begin{equation}\label{eq:complex_derivative}
    \frac{\partial A(B,C)}{\partial B} \Big|_{C} = \left(\frac{\partial A}{\partial x} \frac{\partial C}{\partial y} - \frac{\partial A}{\partial y} \frac{\partial C}{\partial x}\right) / \left(\frac{\partial B}{\partial x} \frac{\partial C}{\partial y} - \frac{\partial B}{\partial y} \frac{\partial C}{\partial x}\right)\,.
\end{equation}

In our program, we calculate the gravitational form-factors from Eq.~\eqref{eq:scalar_charges}, which are complex derivatives of quantities ( e.g. a mass ($m_A$) and moment of inertia ($I_A$) of a NS) taken with respect to external parameters (the value of the scalar field at spatial infinity ($\varphi_0$). The properties of NSs are obtained through numerical integration of the structure equations and depend on the central scalar field ($\varphi_{\mathrm{c}}$) and central pressure ($p_{\mathrm{c}}$). For example for $\alpha_A$ using Eq.~\eqref{eq:complex_derivative} we simply obtain:
\begin{align}
     \alpha_A &= \dfrac{\partial \ln m_A}{\partial \varphi_{0}} \Big|_{\Bar{m}_A} = \left(\frac{\partial \ln m_A}{\partial \varphi_{\mathrm{c}}} \frac{\partial \Bar{m}_A}{\partial p_{\mathrm{c}}} - \frac{\partial \ln m_A}{\partial p_{\mathrm{c}}} \frac{\partial \Bar{m}_A}{\partial \varphi_{\mathrm{c}}}\right) /\nonumber\\ 
     &\qquad\qquad\qquad\qquad\qquad\quad \left(\frac{\partial \varphi_{0}}{\partial \varphi_{\mathrm{c}}} \frac{\partial \Bar{m}_A}{\partial p_{\mathrm{c}}} - \frac{\partial \varphi_{0}}{\partial p_{\mathrm{c}}} \frac{\partial \Bar{m}_A}{\partial \varphi_{\mathrm{c}}}\right),
\end{align}
where every simple derivative is calculated with the machine-precision  due to the AutoDiff.

As a result, the total precision of calculated gravitational form-factors is determined only by the accuracy of the differential equation solver and may be easily enhanced. We use the Julia library \textsc{DifferentialEquations.jl} which allows using AutoDiff on the results of the numerical integration. The numerical integration can take advantage of the AutoDiff itself because the Jacobian of the system can be calculated more accurately and used to integrate the ODE system. We obtain a higher precision if we select a finer adaptive step in the radial variable $r$. For our grids we set the integrator to the relative error of $\delta_{\mathrm{Int, rel}} \sim 10^{-11}$. The obtained relative precision for gravitational form-factors typically lies in $\delta_{\mathrm{rel}}  \sim 10^{-7}-10^{-13}$ range, which is more precise than previous calculations.

\subsection{pre-calculated grids of neutron stars}
\label{App:pre-calculated_grids}

To place limits on the DEF gravity parameters $\{\alpha_0, \beta_0\}$ we need to know the gravitational form factors $\{\alpha_A, \beta_A, k_A\}$ in Eq.~\eqref{eq:scalar_charges} as functions of the mass $m_A$. To approximate these relations we have to calculate grids of NSs. We calculate grids for several EOSs, including MPA1 used in this paper. Each grid contains 4 calculated values: masses $m_A$ and gravitational form factors $\{\alpha_A, \beta_A, k_A\}$ for a range of values of $\{\alpha_0, \beta_0, p_c\}$ with number of points in each axis respectively $\{101,351,121\}$. The grid properties are shown in the Table~\ref{tab:grids}. $\alpha_0$ and $p_c$ are selected to be logarithmically distributed. Whereas, the parameter $\beta_0$ is piece-wise linearly distributed. The grid for $p_c$ is more dense near the common masses of NSs $1.4-2.0 M_\odot$. We also include more points in the region of spontaneous scalarisation with $\beta_0 \in [-5.0, -4.0]$. The resulting calculated gravitational form-factors are saved in files which then can be used in DEF gravity tests, the DDSTG model in particular. 

With the developed DDSTG model we supply grids for 11 different EOSs ranging from soft to stiff. Each EOS satisfies the requirement on a maximal mass in GR: $M_\mathrm{NS,max}^\mathrm{GR} > 2 M_\odot$ \citep{afw+13,fcp+21}. Table~\ref{tab:EOS} presents the selected EOSs with their maximum masses in GR. In our work we use the piece-wise polytropic approximation described in \citet{rea09}.

{\renewcommand{\arraystretch}{1.5}
\begin{table}[t]
    \caption[]{Properties of the pre-calculated grids of NSs.}
    \label{tab:grids}
    \begin{tabular}{llll}
        \hline\hline
        Axis   & Type & \# of points & Range \\
        \hline
        $\alpha_0$ & Log & 101 & $[-10^{-1},-10^{-5}]$ \\
        $\beta_0$ & Lin & 21 & $[-6.0,-5.0]$\\
                         &     & 51 & $[-5.0,-4.0]$ \\
                         &     & 281 & $[-4.0,+10.0]$\\
        $p_c\ (\mathrm{dyn}/\mathrm{cm}^2)$ & Log & 21 & $[10^{34}, 5\times10^{34}]$\\
                         &     & 81 & $[5\times10^{34}, 5\times10^{35}]$\\
                         &     & 21 & $[5\times10^{35}, 10^{36}]$\\                       
            \hline
         \end{tabular}
         \tablefoot{The grids are 3-dimensional and depend on $\{\alpha_0, \beta_0, p_c\}$. Each axis of a grid can be divided into several intervals and be either linearly or logarithmically separated.}
\end{table}
}

\section{PK parameters in DEF gravity}
\label{App:PKs_in_DEF}

In this Appendix we present the expressions for the PK parameters in DEF gravity as functions of the masses of pulsar ($m_\mathrm{p}$) and companion ($m_\mathrm{c}$), and the three Keplerian timing parameters $P_\mathrm{b}$, $e$, and $x$. The formulae for the PK parameters $\{\gamma, k, r, s\}$ can be found in \cite{de92} and \cite{de96new}. For the PK parameters $\{\delta_r, \delta_{\theta}, A, B\}$ there are only more general theory-independent presentations in \cite{dt92}, from which the expressions in DEF gravity, however, can be straightforwardly derived. The total set of PK parameters in DEF gravity reads:
\begin{align}
    &\gamma = \dfrac{e}{n} \dfrac{\Xc \beta_O^2}{1+\alphap\alphac} \left[\Xc(1+\alphap \alphac) + 1 + \kp\alphac \right]\,,\label{eq:ppk_gamma}\\
    &k = \dfrac{3\beta_O^2}{1 - e^2}\left[\dfrac{1 - \frac13\alphap \alphac}{1 + \alphap \alphac} - \dfrac{\Xp \betac \alphap^2 + \Xc \betap \alphac^2}{6(1 + \alphap \alphac)^2}\right]\,,\label{eq:ppk_omdot}\\
    &r = \dfrac{G_* \mc}{c^3}\,,\label{eq:ppk_r}\\
    &s = \dfrac{nx}{\Xc \beta_O} \left[1 +\frac13 \left(\frac{9 - \alphap\alphac}{1+\alphap\alphac} - \Xp\Xc\right)\beta_O^2 \right] \,,\label{eq:ppk_s}\\
    &\delta_r =\dfrac{\beta_O^2}{\mtot^2 (1 + \alphap \alphac)}\left[(3 - \alphap \alphac)\massp^2 \right.\nonumber\\
    & \qquad \left. + (6 - \alphap \alphac - \kp \alphac)\massp \mc + (2 - \alphap \alphac - \kp \alphac)\mc^2 \right] \,,\label{eq:ppk_dr}\\
    &\delta_{\theta} = \dfrac{\beta_O^2}{\mtot^2 (1 + \alphap \alphac)}\left[\left(\frac72 - \frac12\alphap \alphac\right)\massp^2 \right.\nonumber\\
    & \qquad \left. + (6 - \alphap \alphac - \kp \alphac)\massp \mc + (2 - \alphap \alphac - \kp \alphac)\mc^2 \right] \,,\label{eq:ppk_dth}\\
    &A = -\dfrac{\beta_O \Xc}{2\pi\nu_\mathrm{p} (1 -e^2)^{1/2}} \dfrac{\sin \eta}{\sin \lambda}\,,\label{eq:ppk_a}\\
    &B = -\dfrac{\beta_O \Xc}{2\pi\nu_\mathrm{p} (1 -e^2)^{1/2}} \dfrac{\cos i \cos \eta}{\sin \lambda}\,,\label{eq:ppk_b}\\
    &\dot{P}_\mathrm{b} = \dot{P}_\mathrm{b}^{\varphi, \mathrm{mon}} + \dot{P}_\mathrm{b}^{\varphi, \mathrm{dip}} + \dot{P}_\mathrm{b}^{\varphi, \mathrm{quad}} + \dot{P}_\mathrm{b}^{g, \mathrm{quad}} \,,\label{eq:ppk_pbdot}
\end{align}
where we used variables $\beta_O = \left(G_\mathrm{pc} M n/c^3\right)^{1/3}$, $G_\mathrm{pc} = G_* (1 +\alphap\alphac)$, $\mtot = \massp + \mc$, $\Xp = \massp /\mtot$, and $\Xc = \mc / \mtot = 1 - \Xp$. Moreover, $n = 2\pi/P_\mathrm{b}$ is the orbital circular frequency and $\nu_\mathrm{p} = 1 / P$ is the pulsar's rotational frequency. The angle $i$ is the inclination of the orbital plane with respect to the plane of sky, while angles $\lambda$ and $\eta$ are the polar angles of the spin axis. In the special case of alignment between the orbital and spin axes $\lambda = i$ and $\eta = -\pi/2$ forcing $B = 0$.

{\setlength{\tabcolsep}{3.5pt}
\renewcommand{\arraystretch}{1.5}
\begin{table}[t]
      \caption[]{EOSs selected for the calculation of grids of NSs.}
         \label{tab:EOS}
         \begin{tabular}{lcc|lcc}
            \hline\hline
            Name   & $M_{\mathrm{max}}\ (M_\odot)$ & $R_{1.4}$ (km) & Name   & $M_{\mathrm{max}}\ (M_\odot)$ & $R_{1.4}$ (km) \\
            \hline
            SLy  & 2.049 & 11.736 & APR3 & 2.390 & 12.094 \\
            APR4 & 2.213 & 11.428 & WFF1 & 2.133 & 10.414 \\
            WFF2 & 2.198 & 11.159 & ENG  & 2.240 & 12.059 \\
            MPA1 & 2.461 & 12.473 & MS1  & 2.767 & 14.918 \\
            MS1b & 2.776 & 14.583 & H4   & 2.032 & 13.759 \\
            ALF2 & 2.086 & 13.188 & \\
            \hline
         \end{tabular}
         \tablefoot{All EOSs have the maximum mass of a NS in GR $M_\mathrm{NS,max}^\mathrm{GR} > 2 M_\odot$ and taken in piece-wise polytropic approximation from \citet{rea09}. $R_{1.4}$ is the radius of a $1.4M_\odot$ NS.}
\end{table}
}

The expression for $\dot{P}_\mathrm{b}$ is composed from 4 different contributions \citep{de92}
\begin{align}
    &\dot{P}_\mathrm{b}^{\varphi, \mathrm{mon}} = -\frac{3\pi \Xp \Xc}{1 + \alphap \alphac} \,\beta_O^5\, \frac{e^2(1+e^2/4)}{(1-e^2)^{7/2}} \nonumber\\ 
    &\qquad \times \left[\alphap + \alphac + \frac23(\alphap \Xc + \alphac \Xp) + \frac{\betap\alphac + \betac \alphap}{1 + \alphap \alphac} \right]^2,
    \label{eq:ppk_pbdot_mon}\\
    &\dot{P}_\mathrm{b}^{\varphi, \mathrm{dip}} = -\frac{2\pi \Xp \Xc}{1 + \alphap \alphac} \,\beta_O^3\, \frac{1+e^2/2}{(1-e^2)^{5/2}} (\alphap - \alphac)^2  \nonumber\\
    &\qquad -\frac{4\pi \Xp \Xc}{1 + \alphap \alphac} \beta_O^5 \frac{1}{(1-e^2)^{7/2}} \left[ \frac85 \left(1 + \frac{31e^2}{8} + \frac{19e^4}{32} \right)(\alphap - \alphac) \right. \nonumber\\
    &\qquad\left.\times\; (\alphap \Xp + \alphac \Xc)(\Xp - \Xc) +\left(1 + 3e^2 + \frac{3e^4}{8}\right) \right. \nonumber\\
    &\qquad\left. \times\; \frac{(\alphap - \alphac)(\betac \alphap \Xp - \betap \alphac \Xc)}{1 + \alphap \alphac}\right]
    \,,\label{eq:ppk_pbdot_dip}\\
    &\dot{P}_\mathrm{b}^{\varphi, \mathrm{quad}} = -\frac{32\pi \Xp \Xc}{5(1 + \alphap \alphac)}\,\beta_O^5\, \frac{1 + 73e^2/24 + 37e^4/96}{(1-e^2)^{7/2}}\nonumber\\
    &\qquad\times\; (\alphap \Xc + \alphac \Xp)^2 \,, 
    \label{eq:ppk_pbdot_quad}\\
    &\dot{P}_\mathrm{b}^{g, \mathrm{quad}} = -\frac{192\pi \Xp \Xc}{5(1 + \alphap \alphac)}\,\beta_O^5\, \frac{1 + 73e^2/24 + 37e^4/96}{(1-e^2)^{7/2}}
    \,,\label{eq:ppk_pbdot_quadg}
\end{align}
where scalar field $\varphi$ gives rise to a monopolar $\dot{P}_\mathrm{b}^{\varphi, \mathrm{mon}}$, a dipolar $\dot{P}_\mathrm{b}^{\varphi, \mathrm{dip}}$ and a quadrupolar $\dot{P}_\mathrm{b}^{\varphi, \mathrm{quad}}$ contribution. Another quadrupolar term $\dot{P}_\mathrm{b}^{g, \mathrm{quad}}$ is associated with metric $g_{\mu\nu}^\ast$. Following \cite{de92}, we have ignored a term of order $(\alphap - \alphac)^2\beta_O^5$ in Eq.~(\ref{eq:ppk_pbdot_dip}).

\section{DDSTG implementation into \textsc{TEMPO}}
\label{App:DDSTG_implemetation}

As part of this work, we implement a new timing model DDSTG into an independent version of the standard timing software \textsc{TEMPO}\textsuperscript{\ref{url_ddstg}}. The new model is based on the DD \citep{dd86} and DDGR \citep{tw89} models but modified to work within STG theories. In addition to the parameters present in the DDGR model, the DDSTG model utilises four additional parameters, added to the corresponding parameter input file. New parameters and their possible values are presented in Table~\ref{tab:parameters}. 

The first two parameters ``ALPHA0'' and ``BETA0'' correspond to the values of two arbitrary DEF gravity parameters $\{\alpha_0, \beta_0\}$. They can be selected to be zero to recover GR. Otherwise, the applicable range of values is determined by the range of the supplementary data files, containing grids of the calculated gravitational form-factors and masses. The grids are calculated as part of this work and supplied with the model. The supplied grids cover $\alpha_0 \in [-10^{-1},-10^{-5}]$ and $\beta_0 \in [-6,+10]$ in the DEF gravity parameter space.

The third parameter ``EOS'' selects the equation of state to be used to perform the test. In the present work, we generally select the rather stiff EOS MPA1. \textsc{TEMPO} reads the supplied pre-calculated grids of the gravitational form factors and masses corresponding to the selected EOS (see also Table~\ref{tab:EOS}).

The last parameter ``COMP\_TYPE'' is used to select the type of the companion to the pulsar, among a white dwarf (``WD''), a neutron star (``NS'') and a black hole (``BH''). These three compact objects have different properties in terms of gravitational form factors. NSs in DEF gravity can be moderately $|\alpha_\mathrm{NS}| \lesssim |\alpha_0|$ or strongly $|\alpha_\mathrm{NS}| \sim O(1)$ scalarised depending on the mass $m_A$ and the chosen DEF parameters. WDs are not enough compact to significantly deviate from the weak-field approximation $|\alpha_\mathrm{WD}| \lesssim |\alpha_0|$. Thus for WDs we apply $\alpha_\mathrm{WD} = \alpha_0$, $\beta_\mathrm{WD} = \beta_0$.\footnote{Since this is the case for any weakly self-gravitating body, the setting ``WD'' can also be selected for any non-degenerate companion such as a main-sequence star.} In contrast, BHs in DEF gravity fulfil a no-hair theorem and they are completely descalarised $\alpha_\mathrm{BH} = 0$ and $\beta_\mathrm{BH} = 0$. The choice of the companion type plays a significant role for the calculated values of PK parameters.

{\renewcommand{\arraystretch}{1.5}
\begin{table}
      \caption[]{Parameters used in DDSTG timing model in \textsc{TEMPO}.}
         \label{tab:parameters}
         \begin{tabular}{ll}
            \hline\hline
            Name &  Description \\
            \hline
            ALPHA0 & $\alpha_0$ value $\in [-10^{-1},-10^{-5}] \cup \{0\}$ \\
            BETA0 & $\beta_0$ value $\in [-6,+10]$ \\
            EOS & EOS ID, e.g. MPA1\\
            COMP\_TYPE & selected from \{WD,\,NS,\,BH\} \\
            \hline
         \end{tabular}
\end{table}
}

During initialisation, \textsc{TEMPO} reads the parameter file and the file with the TOAs. When the $\{\alpha_0, \beta_0\}$ values and the EOS are provided, it reads the supplementary grids for the selected EOS. The supplied grids are three-dimensional and consist of the masses of NSs $m_A$ and their gravitational form factors $\{\alpha_A, \beta_A, k_A\}$. Two axes are $\{\alpha_0, \beta_0\}$ and the third axis is a fixed range of central pressure ($p_c$) of a NS. 

Bilinear interpolation is performed in the $\{\alpha_0, \beta_0\}$ parameter space for the selected parameters. The result of the interpolation are four single-dimensional arrays for masses and gravitational form factors depending on a range of central pressures. Interpolated one-dimensional arrays are saved in the computer memory and used further without being changed. From this point gravitational form factors are ready to be interpolated in the mass range during the fitting procedure.

\textsc{TEMPO} with selected DDSTG model fits two masses in addition to spin, astrometric and Keplerian parameters. The masses are iteratively changed during the fitting procedure. Only when the masses change \textsc{TEMPO} interpolates the gravitational form factors from the pre-calculated one-dimensional grids. Gravitational form factors are calculated for both the pulsar and the companion and are kept constant for all TOAs within a single iteration. 

\textsc{TEMPO} tries to find the companion mass ($m_{\mathrm{c}}$) and the total mass ($m_{\mathrm{tot}} = m_{\mathrm{p}} + m_{\mathrm{c}}$) corresponding to the best fit. During the fitting, it utilises the derivatives of the timing formula~\eqref{eq:timing_formula} with respect to these two masses. We calculate derivatives separately for each delay in Equations (\ref{eq:delays} -- \ref{eq:delays_end}). We used the approximated expressions for the derivatives of the delays with respect to PK parameters provided by \citet{dd86}. After applying the chain rule for our derivatives we come to expressions depending on the derivatives of PK parameters with respect to the two masses. The model calculates the derivatives of the the PK parameters of DEF gravity (Eqs.~\ref{eq:ppk_gamma} -- \ref{eq:ppk_pbdot}) with respect to the two masses of the timing model, 
\begin{equation}
    \label{eq:derivatives}
    \left(\frac{\partial p^{\mathrm{PK}}_i(m_{\mathrm{c}}, m_{\mathrm{tot}})}{\partial m_{\mathrm{c}}}\right)\Bigg|_{m_{\mathrm{tot}}},
    \quad
    \left(\frac{\partial p^{\mathrm{PK}}_i(m_{\mathrm{c}}, m_{\mathrm{tot}})}{\partial m_{\mathrm{tot}}}\right)\Bigg|_{m_{\mathrm{c}}} \, ,
\end{equation}
which are certainly different to those in GR. Moreover, for proper convergence of the model in highly nonlinear areas the additional corrections to the derivatives related to variation of gravitational form-factors with respect to masses (i.e. $\partial \alpha_A / \partial m_A, \partial \beta_A / \partial m_A$) are required.

The formulae in Eq.~\eqref{eq:derivatives} are large but can be obtained by straightforward differentiation of the PK formulae in Appendix~\ref{App:PKs_in_DEF}. The expressions are incorporated in the model and depend on gravitational form factors, masses and Keplerian parameters which are known on each iteration. Once the derivatives of PK parameters are calculated, the model calculates the derivatives of the timing delays with respect to the two masses and proceeds the iteration. The outcome of the DDSTG model in \textsc{TEMPO} is the same as for the DDGR model. The most important result is the $\chi^2$ value, which is used further to perform tests of gravity.

The DDSTG timing model converges well if the eccentricity of the orbit is not too small and \textsc{TEMPO} is supplied with a reasonable initial guess of the orbital parameters. The model can run into difficulties in the convergence for solutions if those parameters are far away from GR or if the orbit is almost circular. The problem can happen if \textsc{TEMPO} changes masses significantly during the iteration and reaches a point where the Shapiro shape parameter calculated with new masses using Eq.~\eqref{eq:ppk_s} becomes greater than one, i.e.\ $s > 1$. In this case the expression of the Shapiro delay from Eq.~\eqref{eq:delays_shapiro} does not have a physical meaning and \textsc{TEMPO} breaks with an error for TOAs near conjunction, as the argument of the logarithm becomes negative. The convergence can be enforced by applying the special parameter GAIN, which however comes at the cost of increased computational time. A GAIN parameter of less than one forces \textsc{TEMPO} to use smaller steps while iterating. Another way which ensures the convergence is to supply \textsc{TEMPO} with a good initial guess for the masses in DEF gravity. The rough estimate of the masses in DEF gravity can be obtained from the traditional PK method. The combination of these two methods help to converge systems with small eccentricity and Shapiro shape parameter near one ($s \simeq 1$) even in the strongly nonlinear part of the $\{\alpha_0, \beta_0\}$ plane.

\section{Mass-mass diagram near the horn}
\label{App:MM_horn}

   \begin{figure}
   \centering
   \includegraphics[width=9cm]{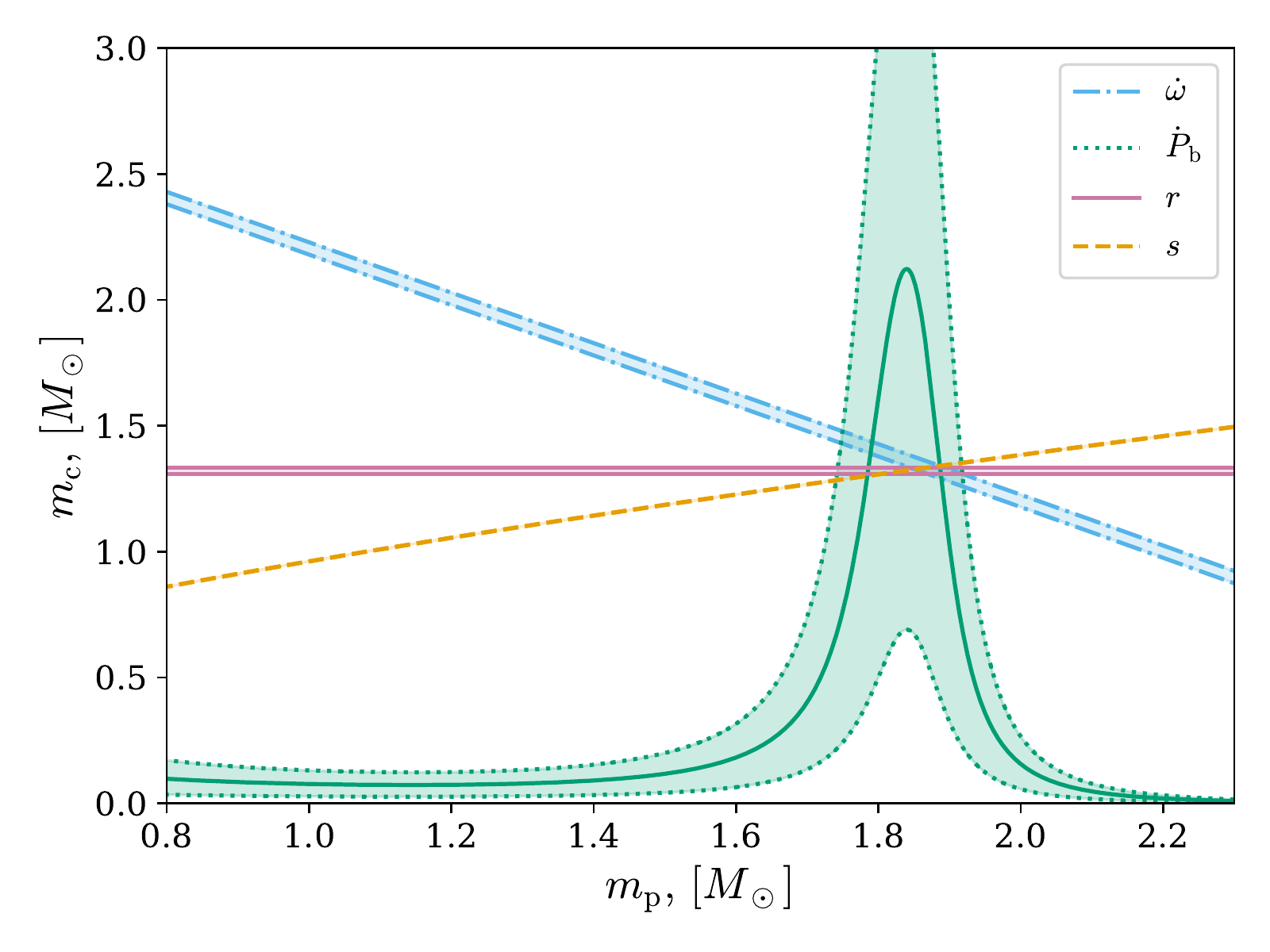}
      \caption{Mass-mass diagram in DEF gravity corresponding to a point near the ``horn'' with $\alpha_0=-10^{-1}$, $\beta_0=-1.9$. The performed test is the same as for Figure~\ref{Fig:J2222_MM_DEF_scalarisation}. The shadowed area is the allowed region at 68 \% CL limit for a corresponding PK parameter. The solid green line corresponds to the observed value of $\dot{P}_\mathrm{b}^\mathrm{int}$.
              }
         \label{Fig:J2222_MM_DEF_hump}
   \end{figure}

Many binary pulsar tests are not sensitive at high $|\alpha_0|$ values and $\beta_0 \sim -2$. In this region near the ``horn'', $\alpha_\mathrm{NS} \approx \alpha_0$ and the overall $\dot{P}_\mathrm{b}$ is close to the GR value, as the dipolar contribution is greatly suppressed. As a consequence, the test is passed. In Figure~\ref{Fig:J2222_MM_DEF_hump} we show a mass-mass diagram in DEF gravity with $\alpha_0 = -0.1, \beta_0=-1.9$. This point in the DEF gravity parameter space passes the test with timing data but is excluded by Cassini experiment ($|\alpha_0| \lesssim 3.4 \times 10^{-3}$). The picture is dramatically different from what we see in the zone of scalarisation in Figure~\ref{Fig:J2222_MM_DEF_scalarisation}.

\section{Proper account of \texorpdfstring{$\dot{P}_\mathrm{b}^{\mathrm{ext}}$}{external PBDOT} uncertainty} \label{App:XPBDOT}

If the observed $\dot{P}_\mathrm{b}$ is measured more precisely than the external contribution $\dot{P}_\mathrm{b}^{\mathrm{ext}}$ we have to take the uncertainty of the external contribution into account. In most situations $\dot{P}_\mathrm{b}^{\mathrm{ext}}$ consists of the Shklovskii and the Galactic contribution. The calculation of the Shklovskii contribution is limited by the uncertainty in the distance and the one in the proper motion. The uncertainty in the estimation of the Galactic contribution is determined by the uncertainty in the distance and our imperfect knowledge of the Galactic gravitational potential. The latter is partly systematic in nature and therefore somewhat more difficult to quantify. If the uncertainty $\Delta\dot{P}_\mathrm{b}^{\mathrm{ext}}$ is the limiting factor we no longer can have $\dot{P}_\mathrm{b}^{\mathrm{ext}}$ fixed on one value for the whole experiment (i.e.\ provide a fixed value for parameter XPBDOT in \textsc{TEMPO}) because its uncertainty can affect the derived limits.

To properly account for the uncertainty in $\dot{P}_\mathrm{b}^{\mathrm{ext}}$ we follow the method described in \citet{sna+02} based on Bayes inference. On the first stage \textsc{TEMPO} calculates $\chi^2$ grid for three independent parameters $\{\alpha_0, \beta_0, \dot{P}_\mathrm{b}^{\mathrm{ext}}\}$. Then the shifted value $\Delta\chi^2 = \chi^2 - \chi^2_\mathrm{min}$ is described by $\chi^2$ distribution with 3 degrees of freedom. The number of degrees of freedom is important when we calculate confidence level limits. The $\Delta\chi^2$ maps as usual to a Bayesian likelihood function,
\begin{equation}
    p(\{t_j\}\,|\,\alpha_0,\beta_0,\dot{P}_\mathrm{b}^{\mathrm{ext}}) = \frac12 e^{-\Delta \chi^2 /2} \, ,
\end{equation}
where $\{t_j\}$ refers to the used timing data. We treat the obtained probability density $p(\{t_j\}\,|\,\alpha_0,\beta_0,\dot{P}_\mathrm{b}^{\mathrm{ext}})$ as the likelihood for Bayes' theorem
\begin{equation}
    p(\alpha_0,\beta_0,\dot{P}_\mathrm{b}^{\mathrm{ext}}\,|\,\{t_j\}) = \frac{p(\{t_j\}\,|\,\alpha_0,\beta_0,\dot{P}_\mathrm{b}^{\mathrm{ext}})}{p(\{t_j\})}p(\alpha_0,\beta_0,\dot{P}_\mathrm{b}^{\mathrm{ext}}) \, ,
\end{equation}
where $p(\alpha_0,\beta_0,\dot{P}_\mathrm{b}^{\mathrm{ext}}\,|\,\{t_j\})$ is the desired joint posterior probability function and $p(\alpha_0,\beta_0,\dot{P}_\mathrm{b}^{\mathrm{ext}})$ is the prior. The Bayesian evidence $p(\{t_j\})$ is obtained from normalisation of posterior probability over the whole grid $\{\alpha_0,\beta_0\}$. 

In the next stage we select the prior $p(\alpha_0, \beta_0, \dot{P}_\mathrm{b}^{\mathrm{ext}})$. We assume no prior information about $\alpha_0$ and $\beta_0$, so their distributions are taken uniformly for $\beta_0 \in (-\infty, +\infty)$ and $\alpha_0 \in (-\infty, 0]$. At this step, we utilise the information about the uncertainties in $\dot{P}_\mathrm{b}^{\mathrm{ext}}$. For example, we can select the prior $p(\dot{P}_\mathrm{b}^{\mathrm{ext}})$ to be normally distributed with the mean measured value $\dot{P}_\mathrm{b}^{\mathrm{ext}}$ and its standard deviation $\Delta\dot{P}_\mathrm{b}^{\mathrm{ext}}$. Or we can use more elaborate probability distribution accounting for our uncertainty in the measurements of distance, proper motion, and Galactic gravitational potential,
\begin{equation}
   p(\alpha_0,\beta_0,\dot{P}_\mathrm{b}^{\mathrm{ext}}) = p(\alpha_0) \times p(\beta_0) \times p(\dot{P}_\mathrm{b}^{\mathrm{ext}}) \, .
\end{equation}

The last step is to marginalise over the $\dot{P}_\mathrm{b}^{\mathrm{ext}}$ variable using the correct joint posterior probability with desired priors
\begin{equation}
    p(\alpha_0,\beta_0\,|\,\{t_j\}) = \int d \dot{P}_\mathrm{b}^{\mathrm{ext}}\, p(\alpha_0,\beta_0,\dot{P}_\mathrm{b}^{\mathrm{ext}}\,|\,\{t_j\}) \, ,
\end{equation}
where the integral is calculated on the grid of $\dot{P}_\mathrm{b}^{\mathrm{ext}}$ parameter. The marginalised probability than can be used to derive limits, for instance going back to $\chi^2$ representation with 2 degrees of freedom.

\end{document}